\documentclass[twocolumn,showpacs,
preprintnumbers,
amsmath,
amssymb,
amsfonts,
aps,
prb,
]{revtex4}

\usepackage{amssymb}
\usepackage{amsmath}
\usepackage[pdftex]{color}
\usepackage[pdftex]{graphicx}
\usepackage{bm}
\usepackage{fancyhdr}
\usepackage[utf8]{inputenc}
\usepackage[francais]{babel}

\def\n{\mathbf{n}}
 \def\r{\mathbf{r}}
 \def\C{\mathcal{C}}
 \newcommand{\ud}{\textrm{d}}
 
\begin{document}
\title{Phase diagram of an extended classical dimer model}
\author{D. Charrier}
\affiliation{Max Planck Institute f\"ur Physik Komplexer System, N\"othnitzer Strasse 38, D-01099 Dresden, Germany}
\affiliation{Laboratoire de Physique Th\'eorique, IRSAMC, Université de Toulouse, CNRS, F-31062 Toulouse, France}
\author{F. Alet}
\affiliation{Laboratoire de Physique Th\'eorique, IRSAMC, Université de Toulouse, CNRS, F-31062 Toulouse, France}

\date{\today}
 
\begin{abstract}
We present an extensive numerical study of the critical behavior of dimer models in three dimensions, focusing on the phase transition between Coulomb and crystalline columnar phases. The case of attractive interactions between parallel dimers on a plaquette was shown to undergo a continuous phase transition with critical exponents close to those of the $O(N)$ \textit{tricritical} universality class, a situation which is not easily captured by conventional field theories. That the dimer model is exactly fine-tuned to a highly symmetric point is a non trivial statement which needs careful numerical investigation. In this paper, we perform an extensive Monte Carlo study of a generalized dimer model with plaquette and cubic interactions and determine its extended phase diagram. We find that when both interactions favor alignment of the dimers, the phase transition is first order, in almost all cases. On the opposite, when interactions compete, the transition becomes continuous, with a critical exponent $\eta 
\sim 0.2$. The existence of a tricritical point between the two regimes is confirmed by simulations on very large size systems and a flowgram method. In addition, we find a highly-degenerate crystalline phase at very low temperature in the frustrated regime which is separated from the columnar phase by a first order transition. 
\end{abstract}

\maketitle

%                                                                                                                                                                                                                                                                                                                                                                                                                                                                           
%%%%%%%%%%%%%%%%%%%%%%%%%%%%%%%%%%%%%%%%%%%%%%%%%%%%%%%%%%%%%%%%%%%%%%%%%%%%%%

\section{\label{sec:Introduction} Introduction}
Originally suggested~\cite{old} as descriptive of adsorption of molecules on a substrate (a motivation that has been renewed in recent experiments~\cite{Blunt}), dimer models have attracted the interest of researchers in various branches of physics, ranging from statistical and condensed matter physics to high-energy physics~\cite{string}. Their distinctive properties essentially result from the close-packing condition, which imposes that on a lattice, each site should be part of one and only one dimer. This strict condition generates strong correlations between degrees of freedom, even when interactions are absent from the system. 

Classical dimer models have been originally studied in statistical mechanics, with the famous result that non-interacting dimer models on planar graphs can be solved exactly using Pfaffians~\cite{60a,60b}. On the square lattice for instance, it was shown using subsequent techniques that dimer correlation functions decay algebraically with distance~\cite{60c}. It was latter shown that dimer models can also be viewed as dual versions of Ising models~\cite{60b} and generate the ground-state manifolds of fully frustrated Ising models~\cite{FF}. For bipartite lattices in three dimensions, dimers can be represented by an effective magnetic field living on the bonds of the lattice. The close-packing condition for the dimers encodes a Gauss law for the magnetic field~\cite{Huse}. Postulating a quadratic dependence on this field of the effective entropic action suggests the existence of {\it dipolar} dimer-dimer correlations. Monte Carlo simulations~\cite{Huse} indeed confirm this picture with a great accuracy, and the corresponding phase of dimers on 3d bipartite lattice is often referred to as a Coulomb phase~\cite{HenleyCoulomb} with this electromagnetic analogy in mind.

In some sense, dimer models form the ``Ising model" of  local constraint, spelling out  their ubiquity in physics. In condensed-matter, dimer models emerged as classical counterparts of quantum dimer models~\cite{QDM} (which properties for particular values of parameters are determined by those of the classical problem), as well as effective models for magnetization plateaus in frustrated magnets~\cite{plateaus}. Dimer models show also close similarities with spin-ice systems~\cite{Isakov} which can also host a Coulomb phase. There the close-packing condition translates into the ``ice rule''.

Intriguing physics take place in classical dimer models when interactions are present. The perhaps simplest case to study is to add local interactions which favor parallel alignment of dimers on a plaquette of the lattice. In two dimensions on bipartite lattices, the system undergoes a phase transition from a columnar phase at low temperature to a disordered critical phase at high temperature~\cite{Alet2d}. The transition is of the Kosterlitz-Thouless type. It is possible to obtain a field theoretical description of the transition in terms of an height model which predicts accurately the behavior of the correlation functions of different observables~\cite{Alet2d,Papanikolaou,Castelnovo}. The situation is much less clear in three dimensions. At high temperature on the cubic lattice, the system is located in the Coulomb phase which is destabilized as the temperature is lowered towards a columnar order of dimers.  Quite surprisingly, the transition between the critical Coulomb phase and the ordered phase is \textit{continuous}~\cite{Alet3d}. Critical exponents estimated from the numerical simulations do not appear to be those of a known ``simple'' universality class, although being very close to those of a tricritical theory.

 A field theoretical description of the transition observed in the $3d$ classical dimer model is not easy. In particular, it cannot be properly addressed in the traditional Ginzburg-Landau formalism since the correlations between degrees of freedom in the disordered phase decay algebraically and not exponentially. Different attempts to provide for a field theoretical description of the critical point have led to a representation in the continuum in terms of two complex matter fields coupled to a non compact $U(1)$ gauge field~\cite{Chalker,Charrier, Chen}, a theory known as the non compact $CP^1$ ($NCCP^1$) theory. The problem is that it is not clear at present if this theory possesses in fact an infra-red fixed point. Efforts to simulate lattice versions of the $NCCP^1$ model either lead to a weakly first-order transition~\cite{Kuklov1}, or to a continuous transition with unconventional critical exponents~\cite{Charrier,Motrunich}. Of course, it is possible that the $NCCP^1$ theory possesses a tricritical point and that the different simulated microscopic models all correspond to the same theory  but flow in the continuum toward different parameter regimes.  
 
 Another possibility is that the microscopic dimer model of Ref.~\onlinecite{Alet3d} sits at a tricritical point, which would be the essentially unique way of reconciling the observed continuous transition with a Ginzburg-Landau approach based on symmetry-breaking. Note that the value of the critical exponents $\alpha \simeq 0.5$, $\nu \simeq 0.5$ and $\eta \simeq 0$~\cite{Alet3d} are suggestive of this scenario as they are close to the ones of a $O(N)$ tricritical theory. This putative tricritical point could be the one of the $NCCP^1$ theory, or of another field theory yet to be specified. It is however unclear why the dimer model should be precisely located at a tricritical point: this usually requires a fine-tuning of parameters, and there is no other parameter than temperature in the original microscopic model~\cite{Alet3d}. \\
 
\begin{figure}[h]
 \begin{center}
\includegraphics*[width=8.5cm]{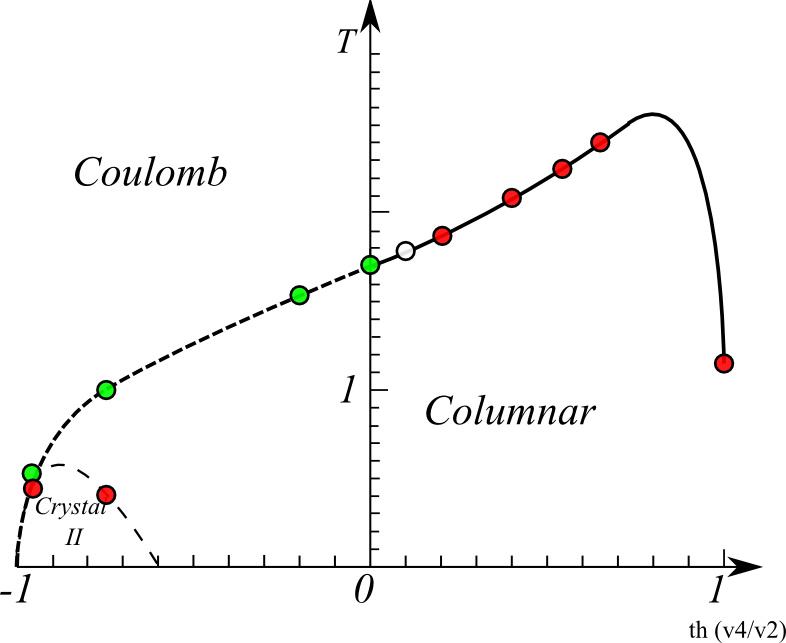}
\caption{Phase diagram of the attractive dimer model with cubic interactions. Dots indicate values of interactions where simulations were performed. Dashed lines and green dots represent second order transitions. Solid lines and red dotes correspond to first order transitions. The white dot represents a point where the nature of the transition is still unclear.}
\label{fig:phasediag}
\end{center}
\end{figure}

 In this paper, we step aside from field theoretical considerations and instead provide new valuable data regarding the critical behavior of 3d classical dimer models. We have carried an extensive Monte Carlo (MC) simulation of a dimer model consisting of a four-dimers interaction on a cubic lattice, in addition to the usual attractive two-dimers plaquette interactions. The cubic interaction corresponds to a coupling between four parallel dimers sitting on the edges of a cube which can be attractive (non frustrated regime) or repulsive (frustrated regime). The system is studied with a worm MC algorithm which allows us to sample  systems of linear size up to $L = 180$. Our results suggest that the phase transition between Coulomb and columnar phases is first order almost everywhere in the non frustrated regime, and second order in the frustrated side - forcing the existence of a tricritical point in between these two regimes. Moreover, we find that the critical exponents in the strongly frustrated regime are different from the ones measured in the absence of the cubic interaction, with $\eta \sim 0.2$ and $\nu \sim 0.6$. In order to rule out  a very weak first order transition for all the range of parameters (which is always possible), we performed a flowgram analysis  which clearly indicates two collapses above and below the tricritical point. Finally, at very low temperature in the strongly frustrated regime, we observe a new crystalline phase resulting from the destabilization of the columnar phase. This phase has a degeneracy growing extensively with the linear system size and is separated from the columnar phase by a first order transition. The final phase diagram that we obtain is presented in Fig.~\ref{fig:phasediag}. We note that similar results were recently obtained by Papanikolaou and Betouras~\cite{Papanikolaou2} in a different deformation of the dimer model. A comparison between the two works will be made in Sec.~\ref{sec:conclusion}.\\
 
 The plan of the paper is the following. In Sec.~\ref{sec:def}, we describe the model, the algorithm and the relevant observables for its study. In Sec.~\ref{sec:nonfrustrated}, we present our results in the non frustrated regime, when both interactions are attractive. In Sec.~\ref{sec:frustated}, we analyze the frustrated regime where the plaquette interaction is attractive but the cubic interaction is repulsive. The final form of the phase diagram is obtained with a flowgram analysis, which we present in Sec.~\ref{sec:flow}. We finally discuss the implications of our results in Sec.~\ref{sec:conclusion}.

 \section{Definitions}
 \label{sec:def}
 \subsection{System}
\begin{figure}[h]
\begin{center}
\includegraphics*[width=\columnwidth]{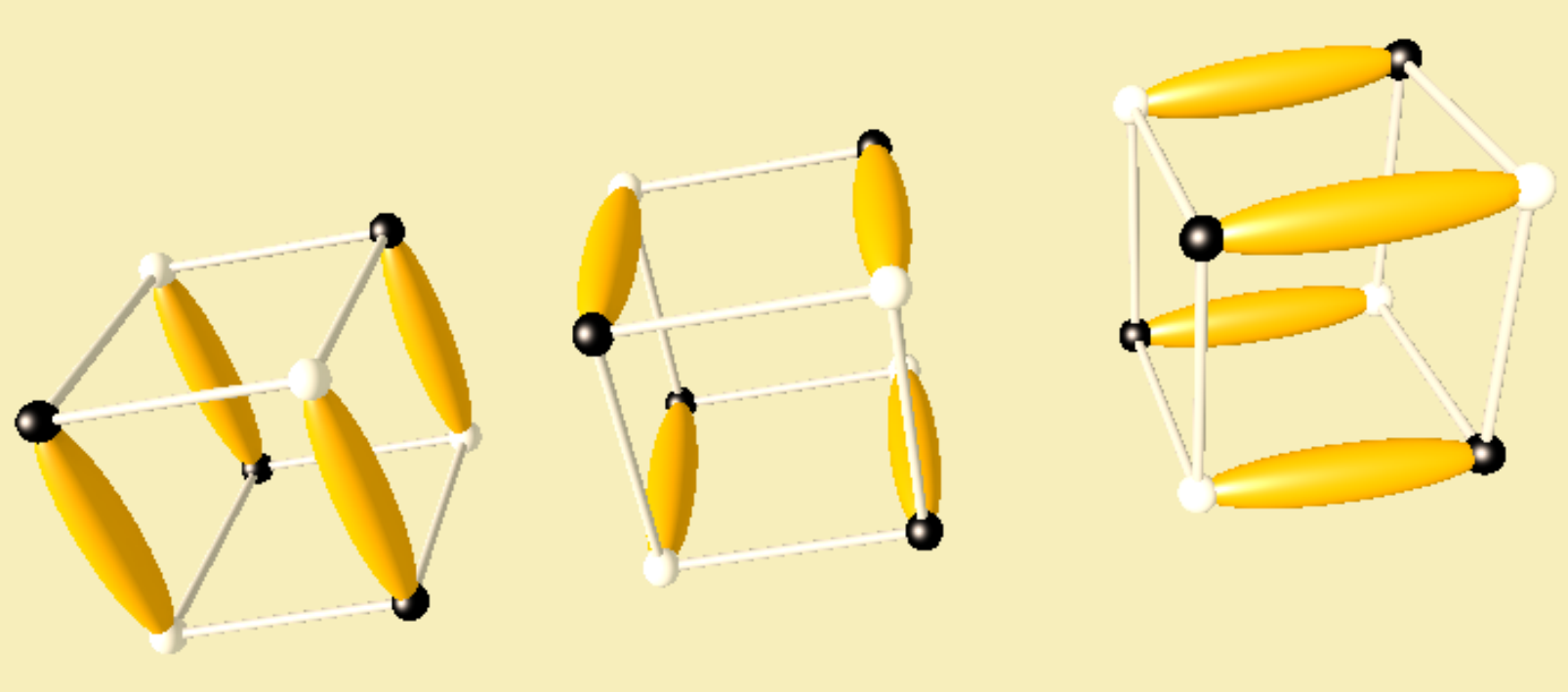}
\caption{The three different possibilities with four parallel dimers parallel on a unit cube. Each pattern contributes $v_4$ to the energy $E_{\C}$ in Eq.~\ref{eq:model}.}
\label{fig:cubes}
\end{center}
\end{figure}

\begin{figure}[h]
\begin{center}
\includegraphics*[width= \columnwidth]{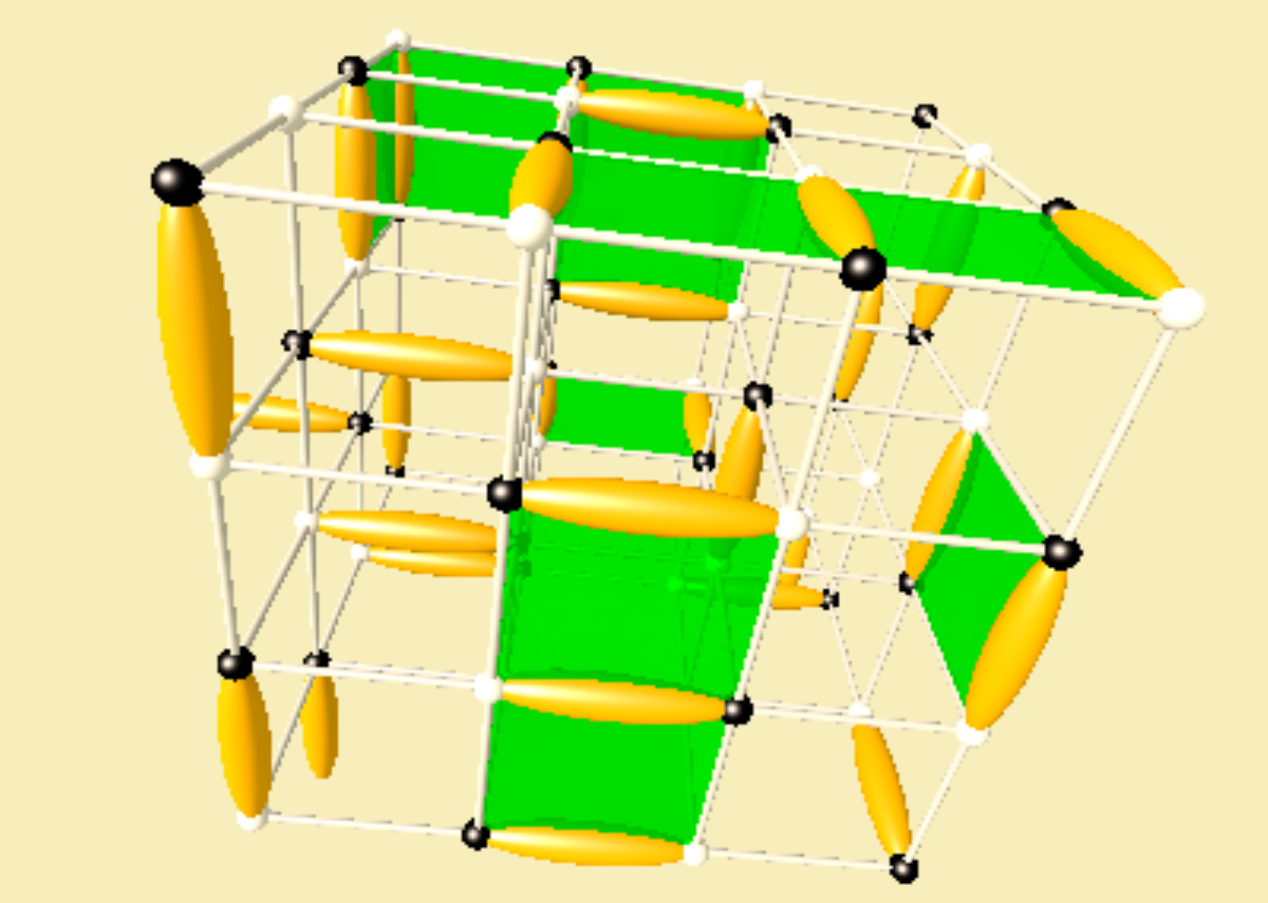}
\includegraphics*[width=\columnwidth]{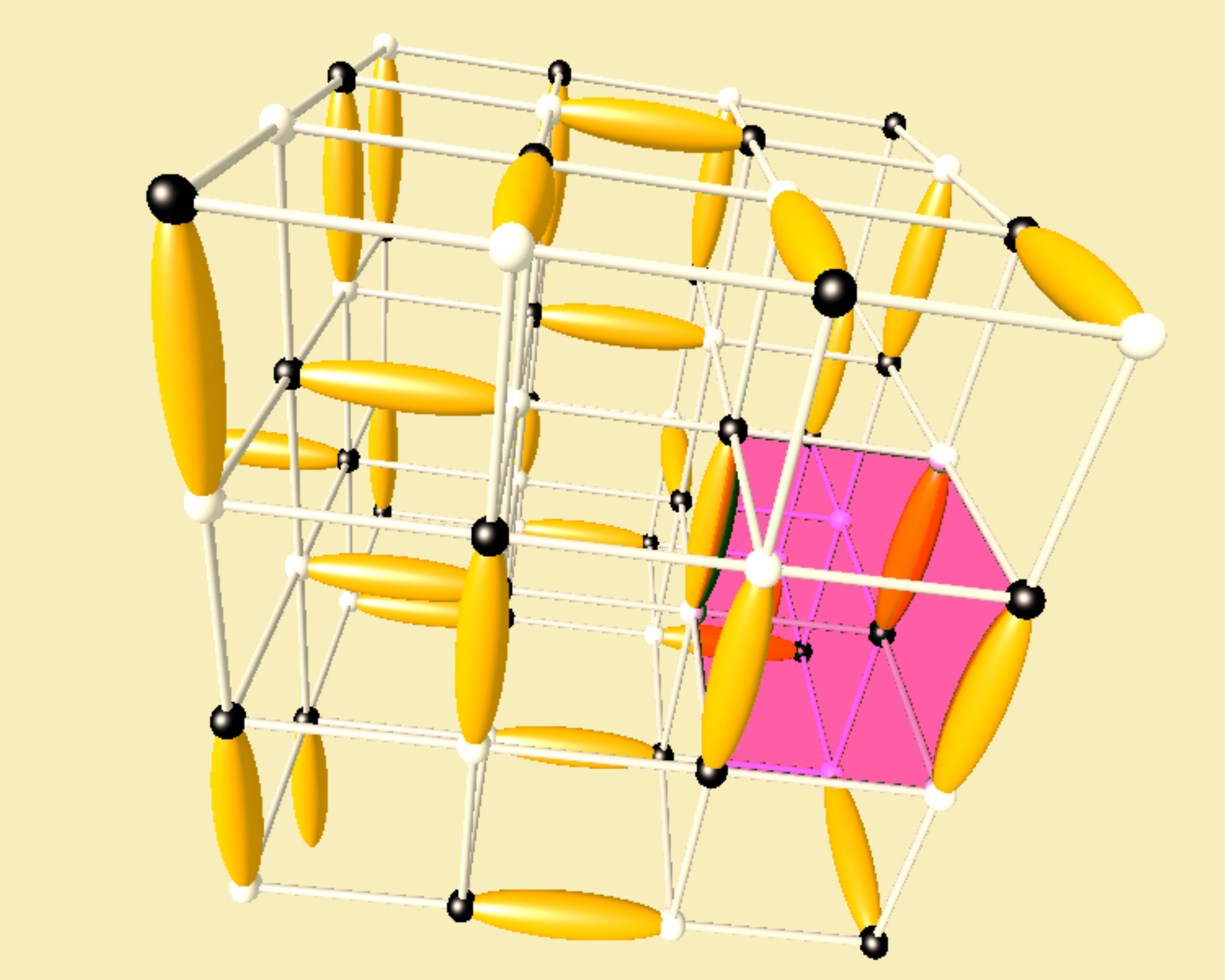}
\caption{Illustration of occurences of the $v_2$ (top) and $v_4$ (bottom) terms of the model defined in Eq.~\ref{eq:model} : interacting plaquettes (top) and cube (bottom) are represented with shaded surfaces. }
\label{fig:3d}
\end{center}
\end{figure}

 The system we consider is a cube of linear dimension $L$ (total number of sites $N=L^3$) covered by hard-core dimers. Only dimer configurations $\C$ obeying the close-packing condition contribute to the partition function. The partition function reads:
\begin{equation}
Z = \sum_{\C} \exp(-\beta E_\C),
\end{equation}
and the energy of an allowed configuration $\C$ is given by:
 \begin{equation}
 \label{eq:model}
 E_\C = v_2( N_{||} + N_{=} +N_{//})+v_4 N_{\rm cubes},
 \end{equation} 
where the first term is proportional to the number of plaquettes in the configuration containing two parallel dimers (referred in the following as ``parallel plaquettes'') and the second counts the number of unit cubes sustaining four parallel dimers (see Fig.~\ref{fig:cubes}). Both occurences of terms in a given dimer configuration are illustrated in Fig.~\ref{fig:3d}. In the remainder of this study, we will restrict ourselves  to the case of attractive plaquette interactions $v_2 < 0$ while cubic interaction $v_4$ can be attractive or repulsive. We investigate the properties of the system as a function of the ratio $x = v_4/v_2$ and temperature $T = 1/\beta$. When $x > 0$, both interactions have the same sign and the system will be said to be non frustrated, as both interactions favor the same columnar ground-states. On the opposite, when $x<0$, the system will be in a frustrated regime ($v_2$ and $v_4$ terms compete). In general, we take $v_2  = -1$ and vary $v_4$. The only exception is the two limits $x \rightarrow \pm \infty$ where we consider the system in absence of plaquette interactions ($v_2 = 0^-$) and take $v_4 = \pm 1$.

We simulate the dimer model by means of a worm Monte Carlo algorithm with a local heat-bath detailed balance condition~\cite{SandvikMoessner}. Compared to a local Metropolis algorithm, autocorrelation times are drastically reduced with a worm algorithm due to the use of non-local moves, which allows us to reach systems of linear size up to $L = 180$.  As we will see in the next section, the ability to simulate very large system sizes is of crucial importance to distinguish between continuous and weakly first-order transitions. We would indeed like to emphasize that for systems like dimer models, or others which contain \textit{a priori} long-range correlations, one should be particularly cautious regarding the issue of finite-size scaling. Here, for the largest system sizes, up to $5. 10^6$ sweeps have been carried (we define one sweep by performing enough worm moves such that on average every site of the lattice is visited). The convergence of simulations is checked by looking at autocorrelation times of the various observables, as obtained from a binning analysis.  
 
 \subsection{Observables}
 \label{sec:observables}
The observables in our system are of three kinds: a first group is made of the thermodynamic quantities such as the average energy, a second type of observables is related to the columnar ordering, and the third type is related to the stiffness of the system (fluctuations of dimer fluxes).

 \subsubsection{Thermodynamic quantities}

A phase transition can generally be detected by monitoring the probability distribution of the energy of the system. For a first order transition, the average energy $\langle E \rangle$ is discontinuous and exhibits a latent heat when the temperature is lowered. For a second order transition, the average energy is continuous but its first derivative, the specific heat per site $C_v/N$, obeys the scaling law:
\begin{equation}
\frac{C_v}{N} = \frac{1}{N}\frac{d\langle E \rangle}{dT}=\frac{\langle E^2\rangle-\langle E\rangle ^2}{T^2 L^3} \simeq C_v^{\rm reg} + A L^{\alpha/\nu}.
\end{equation}
The first term $C_v^{\rm reg}$, corresponding to the regular part of the specific heat at criticality, is often forgotten in fits of numerical data as the divergence of the second term usually dominates (for $\alpha>0$). However, at several points of the phase diagram, we find that this term cannot be neglected as the divergence of the specific heat can turn quite slow. In this situation, one can take $C_v^{\rm reg}$ as either a new fitting parameter or as given by results obtained on small systems where the second term is negligible. In these cases, we take a conservative approach for the error bar on $\alpha / \nu$ and quote a result which encloses all possibilities.

In comparison, for a first order transition, the specific heat diverges like the volume: $C_v/N \propto L^3$. Finally, another mean to distinguish between first and second order transitions is to consider the whole histogram of energy at the transition temperature, as obtained in the Monte Carlo simulation. We expect for a first order phase transition the appearance of double peaks, centered at the average energy of the two co-existing phases. These peaks will appear only for samples with size above (or close  to) the correlation lenght at the transition, and should keep being separated when increasing system size. For a second order phase transition, the histogram should contain a unique peak. 

\subsubsection{Columnar order parameter}
The local columnar order parameter $\mathbf{m}(\r)$ is defined with respect to the dimer occupation number at each site $\n(\r)$:
\begin{equation}
\mathbf{m}(\r) = (-1)^\r \n(\r).
\end{equation}
The global order parameter reads $\mathbf{C} = \frac{2}{L^3} \sum_\r \mathbf{m}(\r) $ and is normalized such that the six columnar states correspond to: $\mathbf{C} = \{ \pm 1, 0,0 \} , \{ 0 , \pm 1, 0 \} , \{ 0,0,\pm 1 \}$ and $C = \langle | \mathbf{C} | \rangle = 1$. One also considers the corresponding susceptibility $\chi$. In particular, for a second order transition:
\begin{equation}
\chi/N = \frac{\langle \mathbf{C}^2\rangle - \langle |\mathbf{C}|\rangle^2}{L^3} \propto L^{2-\eta},
\end{equation}
while for a first order transition $\chi/N \propto L^3$. Finally, the Binder cumulant:
\begin{equation}
B = \frac{\langle \mathbf{C}^4\rangle}{\langle \mathbf{C}^2\rangle^2}.
\end{equation}
is a scale-invariant quantity in the case of a continuous transition, and should thus exhibit a crossing point at criticality as a function of the system size. Moreover, the finite size-scaling (FSS) of its derivative with respect to the temperature: 
\begin{equation}
\frac{\ud B}{\ud T} \propto L^{1/\nu},
\end{equation}
gives a direct access to the critical exponent $\nu$.
 \subsubsection{Stiffness}
The (inverse) stiffness encodes fluctuations of dimer fluxes across a plane~\cite{Huse,Alet3d}:
\begin{equation}
K^{-1} = \sum_{\alpha = x,y,z} \frac{\langle \phi_\alpha^2\rangle}{3L}
\end{equation}
where the flux $\phi_\alpha$ is the algebraic number of dimers crossing a plane
perpendicular to the unit vector $\hat{\alpha}$ . Given a lattice direction, the contribution to the flux is $+1$ for a dimer going
from one sublattice to the other and $-1$ for the reverse situation. The stiffness is finite in the Coulomb phase, reflecting the presence of dipolar correlations between dimers. On the other hand, the columnar phase is robust to insertion of fluxes and $K^{-1}$ vanishes exponentially with system size. At a second order phase transition, the quantity $L K^{-1}$ should be scale invariant~\cite{Alet3d} and the scaling of its derivative:
\begin{equation}
L.\frac{ dK^{-1}}{dT} \propto L^{1/\nu}
\end{equation}
provides another access from the high temperature side to the exponent $\nu$. In general, the error bars that we quote on exponents include at the same time errors due to the fitting procedure (which we measure by considering stability of fits with exclusion of a few data points), errors from the determination of critical temperature as well as statistical errors.

\section{Non frustrated side: $v_2 < 0$, $v_4 < 0$.}
\label{sec:nonfrustrated}
We start the discussions of our numerical results on the non-frustrated side $v_4 < 0$ and $v_2 < 0$ of the phase diagram. When both interactions are attractive, we expect to find the same phases as in the simple attractive plaquette model: a six-fold degenerate columnar phase at low temperature and a Coulomb phase with dipolar correlations at high temperature. 

\begin{figure}[h]
 \begin{center}
\includegraphics*[width=7cm]{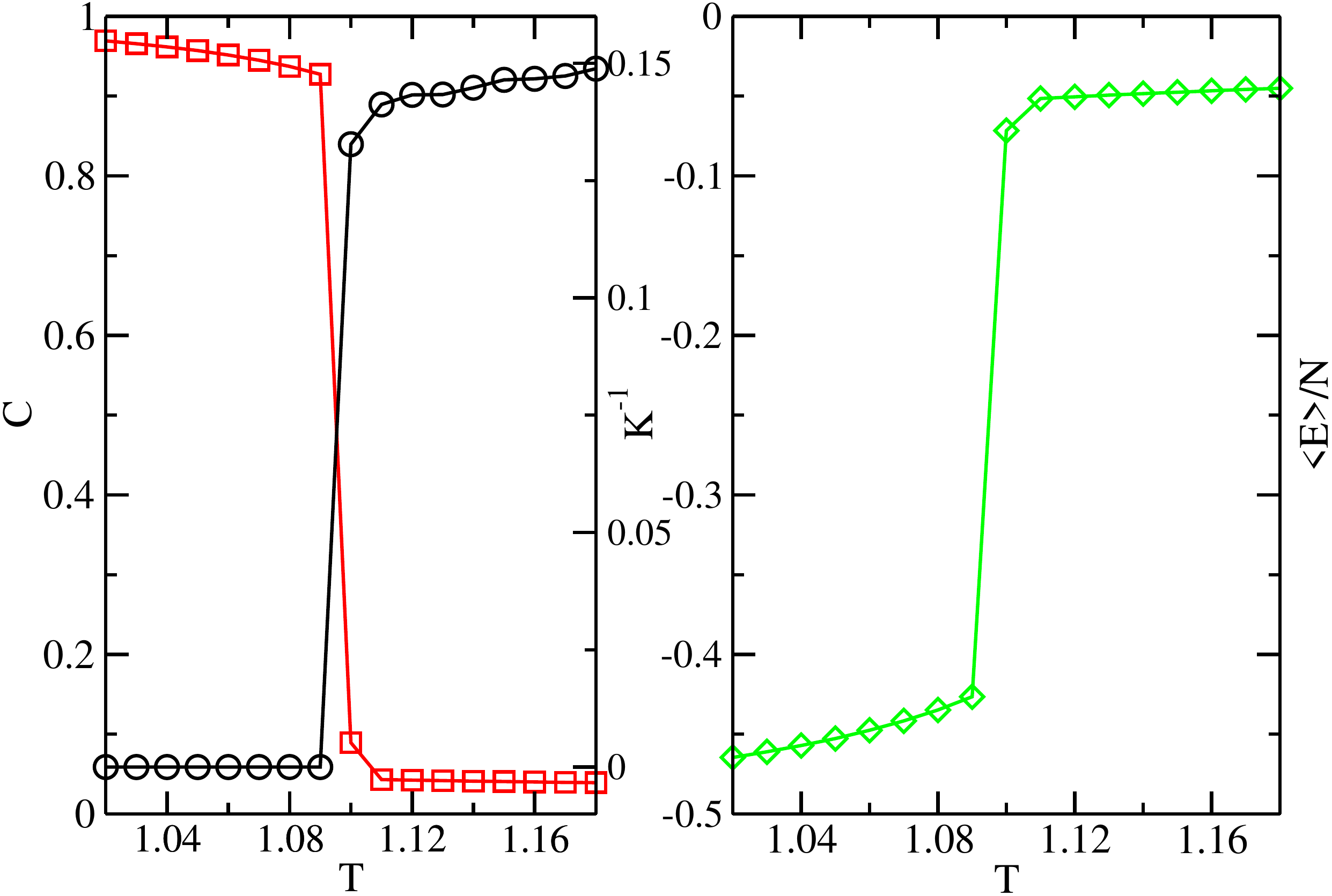}
\caption{Left: Evolution of the columnar order parameter $C$ and inverse stiffness $K^{-1}$ as a function of temperature for $v_2=0$ and $v_4=-1$. Right: Temperature dependence of the average energy per site $\langle E \rangle/N$. Here L = 16.}
\label{fig:extremeNF}
\end{center}
\end{figure}

At first, let us consider the extreme case where only attractive cubic interactions are present ($v_2=0$ or $\tanh (v_4/v_2 ) = 1$ in the phase diagram of Fig.~\ref{fig:phasediag}). The evolution of the energy, columnar order parameter $C$ and inverse stiffness $K^{-1}$ for small system sizes is presented on Figure \ref{fig:extremeNF}. As expected, the columnar order is non-zero only at low temperature where the inverse stiffness vanishes. But in constrast with the attractive plaquette model, both quantities exhibit a strong discontinuity at the transition temperature $T_c \sim 1.1$. In fact, the energy also displays such a jump, characteristic of a latent heat. This shows undoubtedly a first order transition.   

\begin{figure}[h]
 \begin{center}
\includegraphics*[width=8cm]{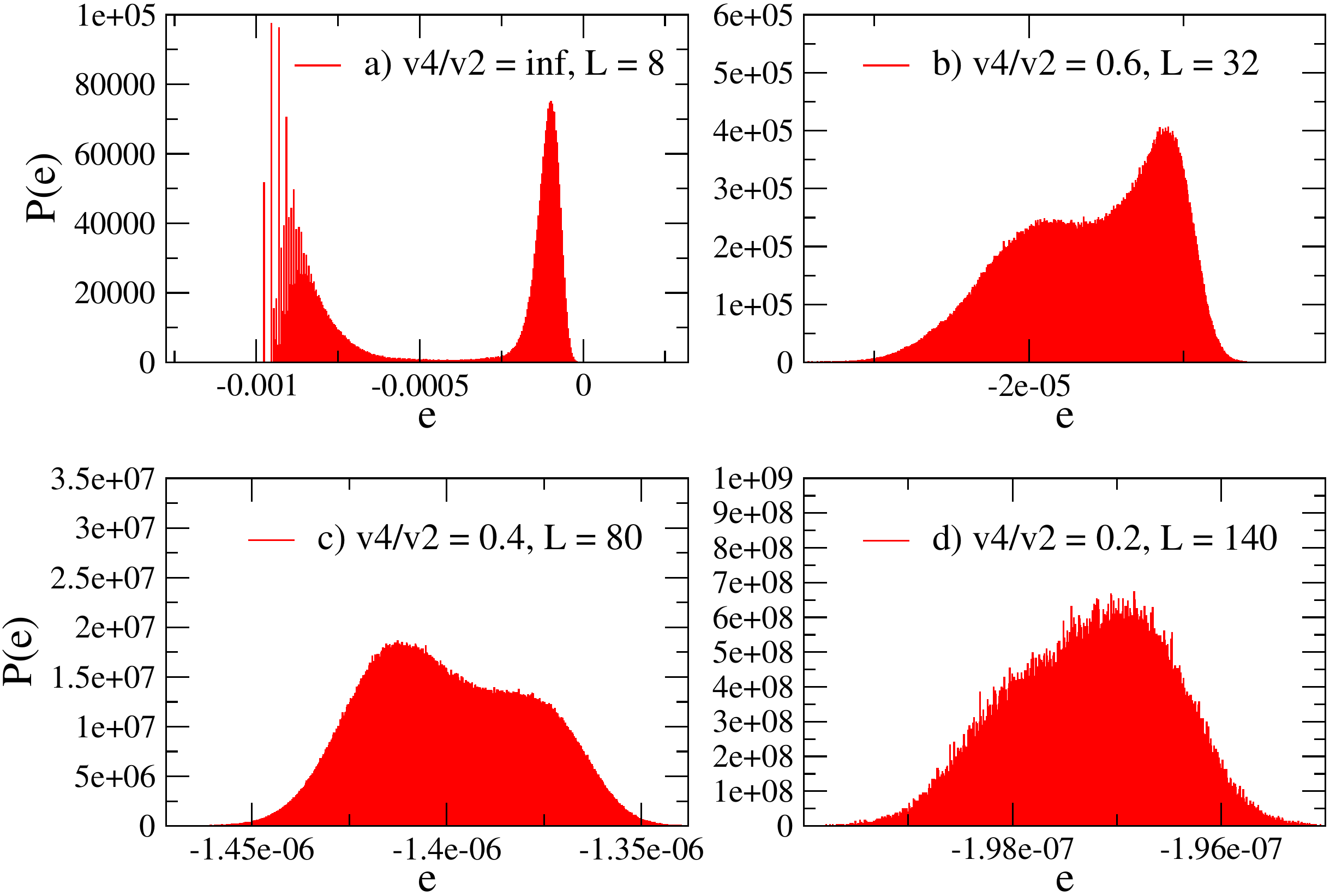}
\caption{Energy per site probability $P(e)$ as a function  of energy per site $e=E/N$, for different ratios $v_4/v_2>0$ at the critical temperature separating columnar and Coulomb phases. The size $L$ corresponds to the minimal length for which the histograms start to display a double peak distribution. For $v_4/v_2 = 0.2$, a deformed single peak is observed in the distribution for $L = 140$, the maximal sample size simulated for this parameter.}
\label{fig:histos}
\end{center}
\end{figure}

To settle definitively the nature of the transition as well as to benchmark the method, we also study the histogram of energy observed during the simulation, which corresponds to the probability distribution of the energy $P(E)$. The appearance of a double peak distribution at the critical temperature is a typical sign of a first order transition. For $v_4/v_2 = \infty$, we can easily detect this double peak for system sizes as small as $L = 8$ (see Fig.~\ref{fig:histos}). 

We now introduce a small attractive plaquette interaction $v_2$ and repeat the procedure by tracking the temperature at which columnar order sets in and inverse stiffness vanishes. We observe that the nature of the transition remains discontinuous but that the correlation length grows as the ratio $v_4/v_2$ is decreased: for $v_4/v_2 = 0.8$ we find double peaks only at sizes $L \geq 16$, for $v_4/v_2 = 0.6$ at sizes $L \geq 32$ and for $v_4/v_2 = 0.4$ at $ L \geq 80$ (see Fig.~\ref{fig:histos}). Finally, for systems close to the pure plaquette model $v_4/v_2 = 0$, it becomes extremely difficult to distinguish double peaks in the energy distribution. In fact, the histogram for $v_4/v_2 = 0.2$ shows a slightly deformed single peak for $L = 140$. We were not able to see any sign of double peaks for $v4/v2 = 0.1$ up to $L=140$.  

Another possibility to discern a discontinuous transition is to measure the critical scaling of the maximum of the susceptibility and the specific heat per site. At a first order transition, both quantities should diverge like the volume $L^3$. In terms of the critical exponents introduced in section \ref{sec:observables}, this would correspond to effective critical exponents $\left.\frac{\alpha}{\nu}\right|_{\rm eff}= 3$ and $\eta_{\rm eff} = -1$. We have determined the scaling laws of these two quantities for several values of the ratio $v_4/v_2$ (see table \ref{tab:tableau1}) and find that when $v_4/v_2$ is large, the exponents agrees with the first order values. As we approach $v_4/v_2 = 0$, the scalings of $C_v$ and $\chi$ get smoother and the exponents closer to the values of Ref.~\onlinecite{Alet3d} $\left.\frac{\alpha}{\nu}\right|_{\rm eff} \sim 1$ and $\eta_{\rm eff} \sim 0$. At this point, it is not possible to conclude on the order of the transition at $v_4/v_2 = 0.1$. The transition can be either continuous or very weakly first order. 

\begin{center}
\begin{table}[h]
\begin{tabular}{|c|c|c|c|c|c|}
\hline 
$v_4/v_2$ & $T_c$ &$ L_{\rm min}$ & $ L_{\rm max}$ &$\left.\frac{\alpha}{\nu}\right|_{\rm eff}$ & $\eta_{\rm eff}$ \\
\hline 
0.6 & $2.225$&$ 16 $& $32$ &$ 3 $ & $-1.00$\\
\hline
0.4 & $2.033$&$ 56 $ & $80$ & $ 2.9$& $ -0.90$ \\
\hline
0.2 & $1.851$&$ 64$ & $140$ & $2$ &$-0.40$ \\
\hline
0.1 & $ 1.7625 $ &$64 $&$140$&$1.4$& $-0.25$ \\
\hline
0 & $ 1.765$&$56 $ & $96$&$ 1.11(5)$ & $ -0.02(5) $ \\
\hline
\end{tabular}
\caption{Critical temperature and effective critical exponents for different non frustrating coupling ratios. For each ratio, the exponents have been measured by a FSS analysis using $L_{\rm min}$ as the minimal size. The values at $v_4/v_2=0$ are taken from Ref.~\onlinecite{Alet3d}.}
\label{tab:tableau1} 
\end{table}
\end{center}

\section{Frustrated side: $v_2 < 0$ $v_4 > 0$}
 \label{sec:frustated}
 
 We now turn to the analysis of the system where plaquette interactions are attractive but cubic interactions are repulsive. As the two interactions compete with each other, we can expect, at least at low temperature and in the regime $v_4/v_2 \ll -1$, that new phases may appear. 

\begin{figure}[h]
 \begin{center}
\includegraphics*[width=7cm]{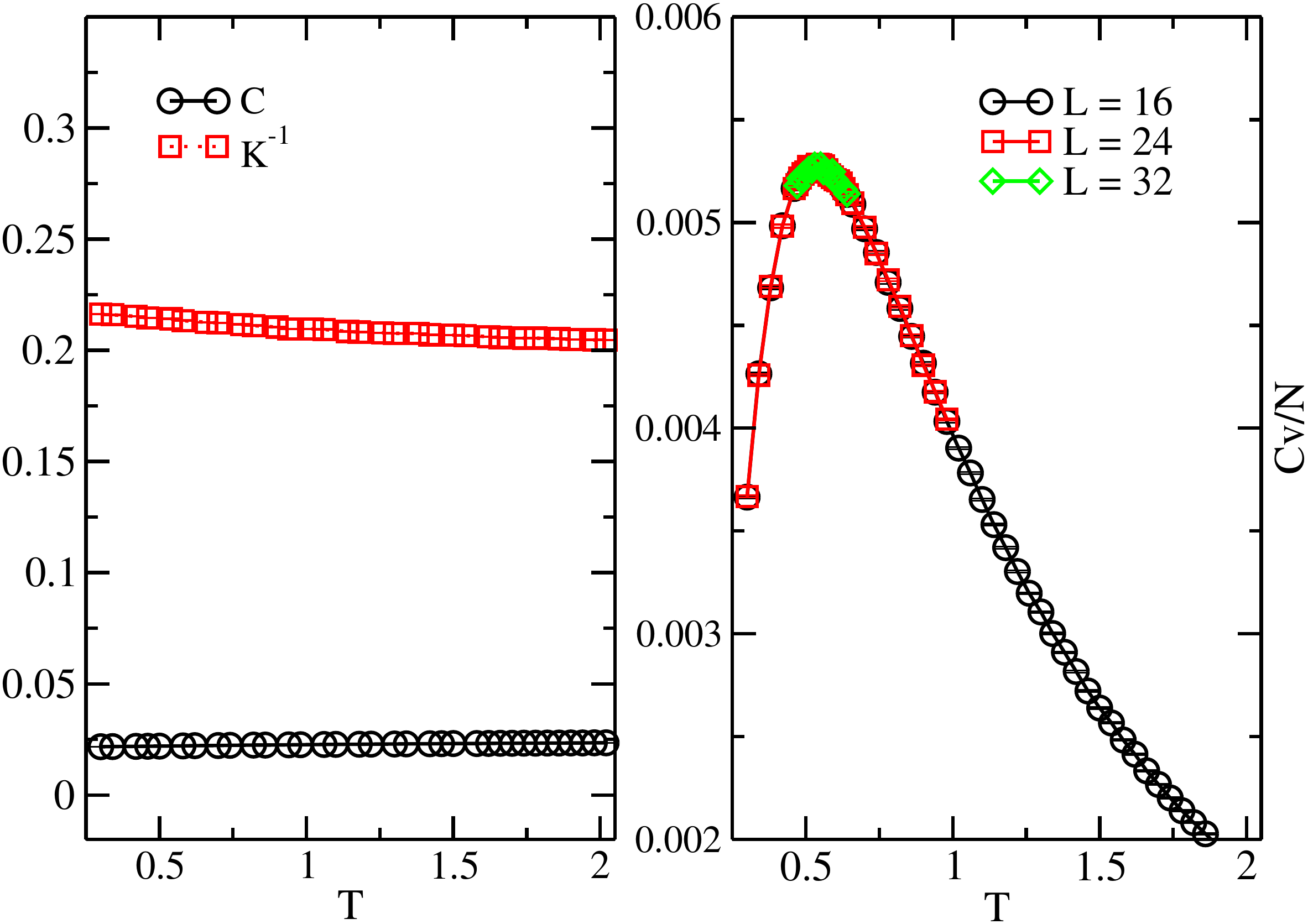}
\caption{For $v_4 = 1$ and $v_2 = 0$. Left: Evolution of the stifness $K^{-1}$ and columnar order $C$ as a function of temperature for $L=16$. The non-zero value reached by $C$ is due to the finite size of the sample. Right: The height of the peak of the specific heat per site does not display any dependence on system size.}
\label{fig:extremeF}
\end{center}
\end{figure}

Let us again first discuss the extremal case where only cubic interactions are present ($v_2 = 0$). We observe that the columnar order parameter vanishes for all temperature while the inverse stiffness $K^{-1}$ remains always non-zero and finite (see Fig.~\ref{fig:extremeF} left). Moreover, although the specific heat displays a maximum (see Fig.~\ref{fig:extremeF} right), it does not display any dependence on the size of the system and cannot be related to a critical phenomena. Thus, we obtain the surprising result that the system is always disordered when $v_4 > 0$ and $v_2 = 0$. In other words, the Coulomb phase can accomodate for having no cubes occupied by four dimers (as in Fig.~\ref{fig:cubes}), even down to very low temperatures.  Even more, the Coulomb phase appears to be strenghten in this situation as the inverse stiffness increases slightly as temperature is lowered. 
   
\begin{figure}[h]
 \begin{center}
\includegraphics*[width=8cm]{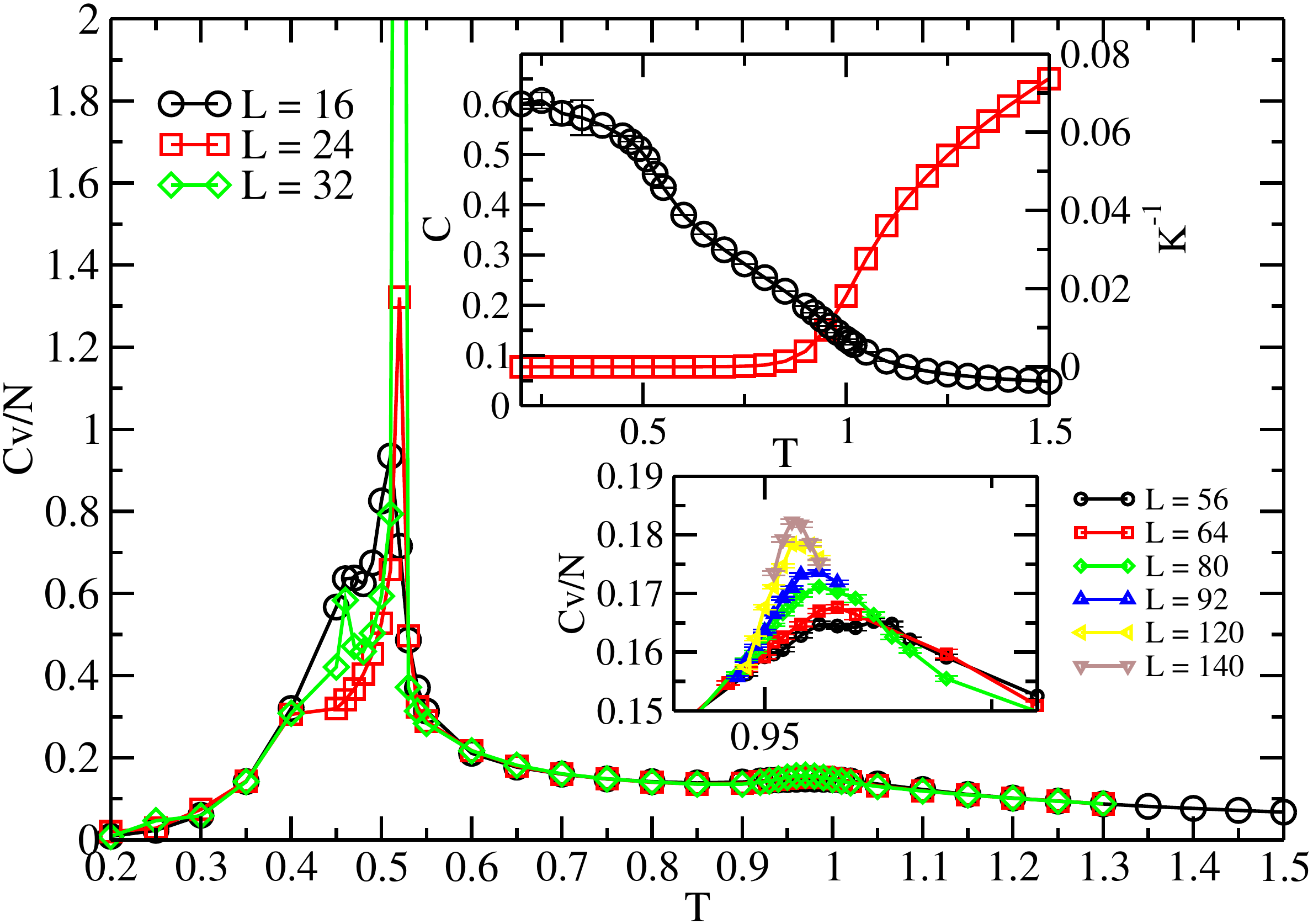}
\caption{Evolution of the specific heat per site as a function of temperature for $v_4 /v_2= -1$. The first peak at $T_{c_1} \sim 0.5$ is associated with the crystallisation of the system in an ordered phase different from the columnar phase, which can be readily seen by the evolution of the columnar order parameter (see upper inset). The second peak, at $T_{c_2} \sim 0.95$, delimitates a region with non-zero columnar order from  a region with non-zero inverse stiffness. Upper inset: evolution of the inverse stiffness $K^{-1}$ and of the columnar order parameter $C$ for $L = 16$. Lower inset: zoom of the specific heat close to $ T_{c_2}$ . }
\label{fig:result1}
\end{center}
\end{figure}

As soon as a finite attractive coupling $v_2$ is introduced, we find that the specific heat per site displays two peaks as a function of temperature: one at a lower temperature $T_{c_1}$ which is strongly diverging with the system size and another at an upper temperature $T_{c_2}$ which diverges very slowly. Fig~\ref{fig:result1} and its insets display results at $v_4/v_2=-1$, which are typical of what we observe in the frustrated regime. The first peak of the specific heat is associated with the freezing of the columnar order parameter to a value smaller than $1$ at $T < T_{c_1}$ , the second to the appearance of the Coulomb phase for $T>T_{c_2}$ (see upper inset of figure \ref{fig:result1}). In the next two sections, we will detail the nature of these two phase transitions and of the phase separating them. We note that as the ratio $v_4/v_2 \rightarrow - \infty$, the two critical temperatures get closer such that it becomes more difficult to detect the peak at $T_{c2}$. 

 \subsection{Low temperature phase transition $T_{c_1}$}
 
\begin{figure}[h]
 \begin{center}
\includegraphics*[width=8cm]{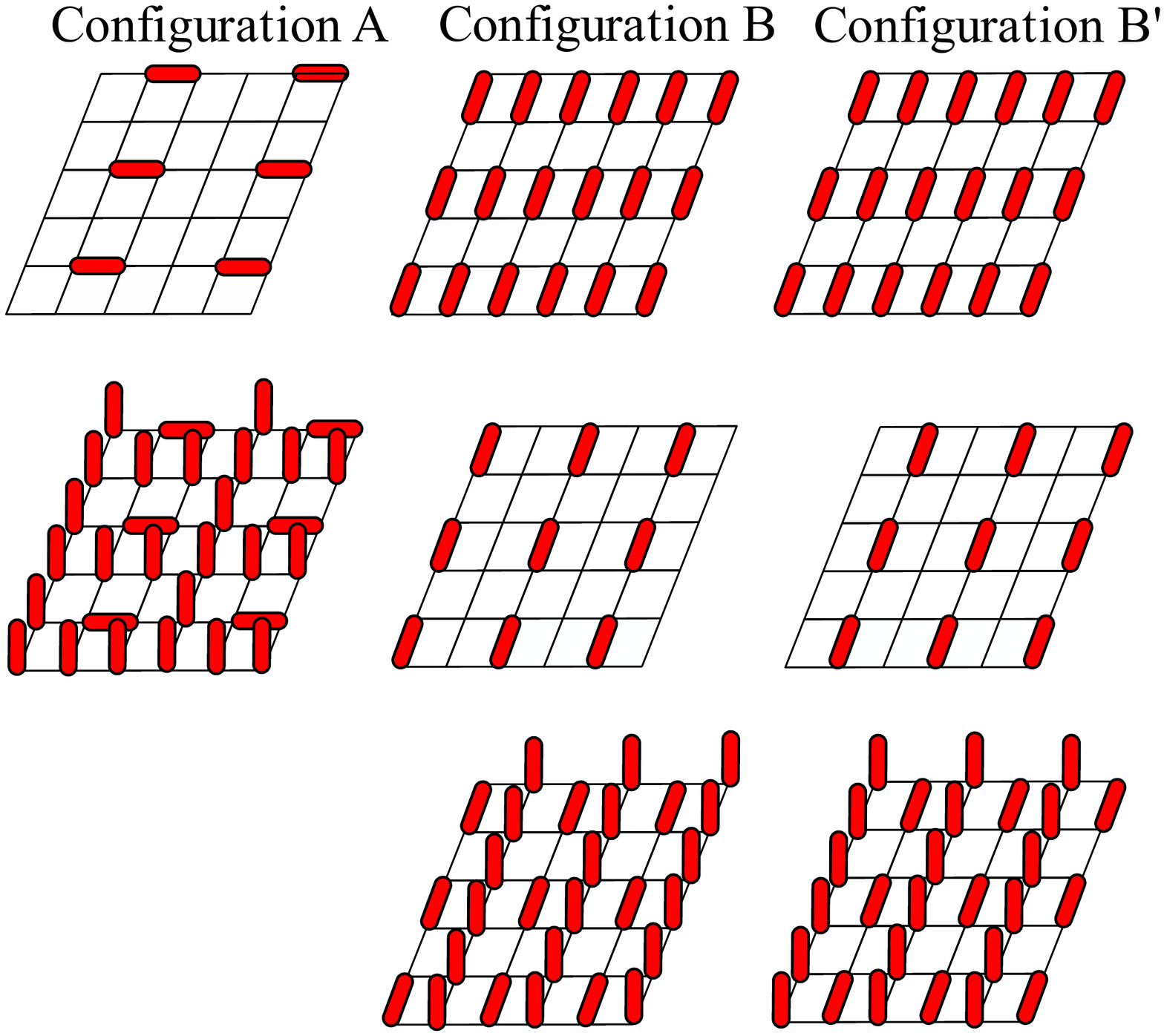}
\vspace{-4cm}
\caption{Three configurations on the $L=6$ cube which maximize the number of parallel plaquettes without having any parallel cubes. Configuration $A$ is built out of a unit pattern containing two planes which repeats itself three times. This unit pattern possesses $42$ parallel plaquettes, leading to a number of $7/12$ parallel plaquettes per unit site.  Configurations $B$ and $B'$ both consist of a unit pattern of three planes repeating itself twice. The unit pattern possesses $63$ parallel plaquettes, leading also to a number of $7/12$ parallel plaquettes per unit site. Configuration $B'$ is obtained from configuration $B$ by shifting the dimer pattern on the two lowest planes by one unit cell on the right direction. Mixing $B$ and $B'$ patterns, one can create several dimer configurations having the same energy.}
\label{fig:packing}
\end{center}
\end{figure}

 We now discuss the nature of the phase below the lower transition by first considering the evolution of the $T=0$ ground-states as a function of the frustration ratio $v_4/v_2$. For large positive values of $v_4$ (but still with $v_2=-1$), we expect the columnar ground-states to become unstable as they maximize the number of parallel cubes. In this limit, we must find dimer configurations which have exactly zero parallel cubes but that can nonetheless support a maximal number of parallel plaquettes. We find that there are several configurations  (in fact an exponential number) that satisfy this frustrating condition. For instance, we present in Fig.~\ref{fig:packing} on the example of a $L=6$ cube three configurations satisfying these two constraints. Configuration $A$ is made of a unit pattern consisting of two planes repeating each other. The unit pattern of configurations $B$ and $B'$ posseses three planes. Configurations $B$ and $B'$ are simply related by a unit translation of the two bottom planes. If we now consider larger systems, it is easy to see that we can use the same plane patterns at will, by randomly choosing $B$ or $B'$ every three planes and this, without creating parallel cubes. Therefore the degeneracy of the ground-state is at least growing like $2^{L/3}$, that is exponentially with the linear system size. It is possible that by forming other cost-free defects within a plane, the degeneracy is even higher $\propto e^{a L^2}$, but we have not made any further investigations in this direction.
 
\begin{figure}[h]
 \begin{center}
\includegraphics*[width=7cm]{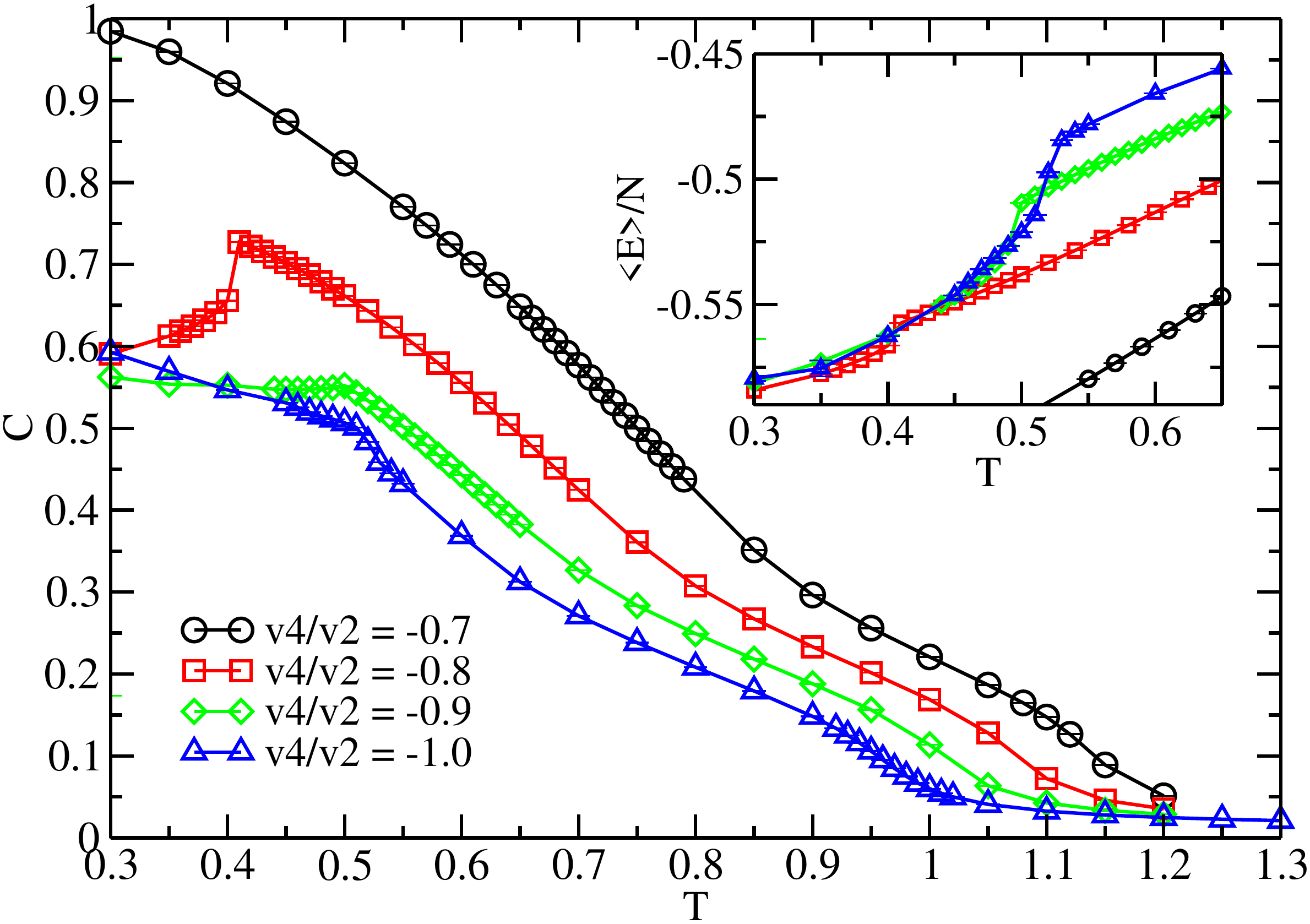}
\includegraphics*[width=6.5cm]{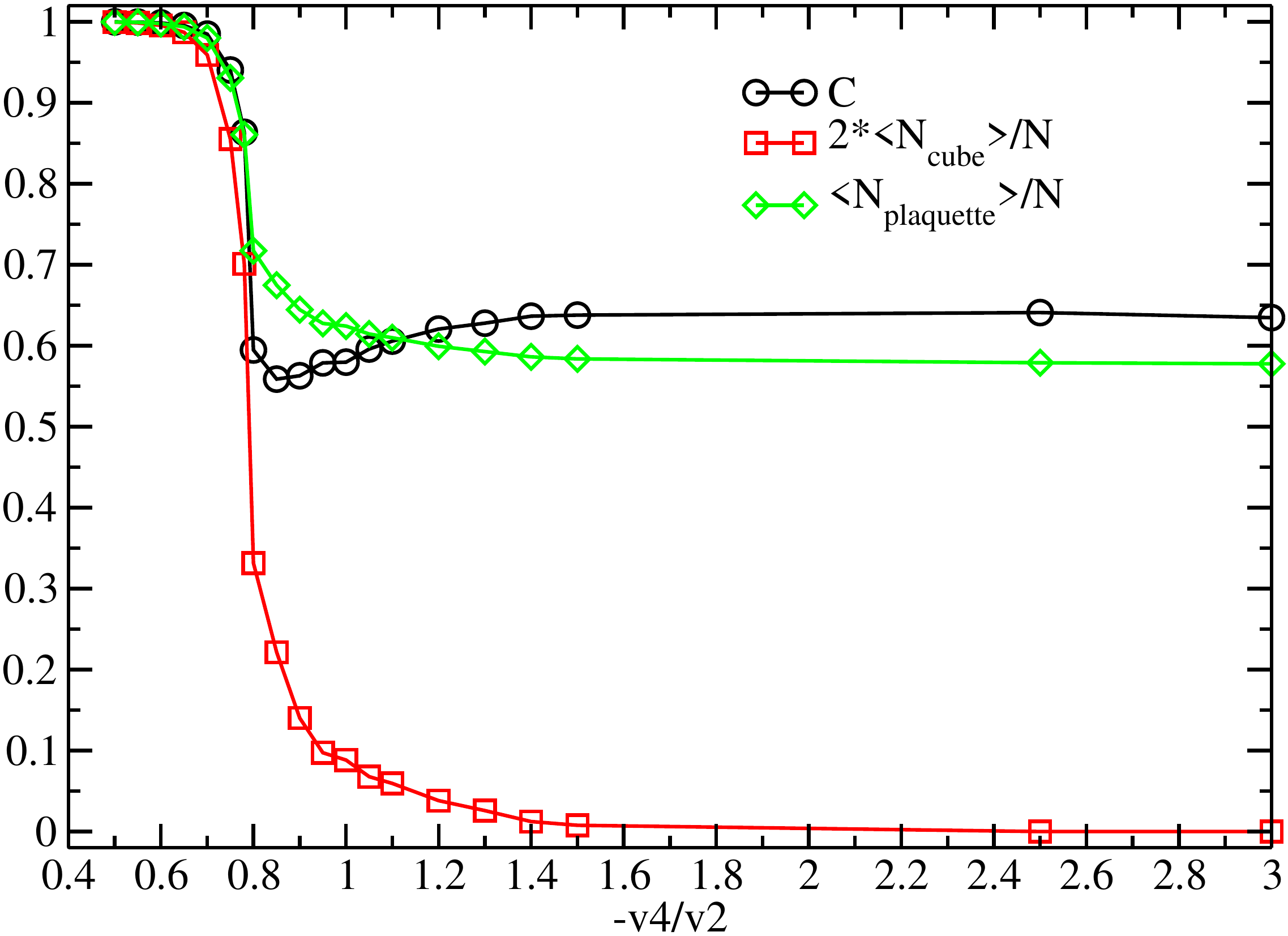}
\caption{Top: Evolution as a function of temperature of the columnar order parameter $C$ for different ratios $v_4/v_2$ at $L = 32$. For $v_4/v_2 < -0.8$, $C$ does not converge towards $1$ at $T=0$. Inset: Evolution of the average energy per site. A discontinuity corresponding to the phase transition can easily be deteted. Bottom: Evolution of $C$, $\langle N_{\rm plaquette} \rangle/N$ and $\langle N_{\rm cube} \rangle/N$ as a function of $v_4/v_2$ for $T = 0.3$ and $L = 32$.}
\label{fig:crystal}
\end{center}
\end{figure}

For what thermodynamics is concerned, it is easy to check that all such configurations with no parallel cubes have on average $7/36$ plaquettes that are parallel. We refer to the corresponding phase as Crystal II. While it may be possible to define correctly an order parameter for this crystal (and this in spite of the high ground-state degeneracy), we simply concentrate on a simple comparison between the energy of the Crystal II ground-states with those of the columnar phase. A columnar ground-state satisfies $1/3$ of the possible parallel plaquettes, and one parallel cube out of two. Since the number of plaquettes (cubes) is three times (equal to) the number of lattice sites, we obtain:
\begin{eqnarray}
E_{\rm Crystal \, II} / L^3 &=& E_{\rm columnar} / L^3\nonumber \\
\Leftrightarrow \frac{7}{12} v_2 &=& v_2 - \frac{v_4}{2} \\
\Leftrightarrow \frac{v_4}{v_2} &=& -\frac{10}{12}. \nonumber 
\end{eqnarray}
That is, the Crystal II phase should be favored as soon as $v_4/v_2 < -10/12\simeq 0.833$. We now compare this simple estimate with numerical simulations. Considering the evolution of the columnar order parameter as a function of $T$ for different ratios $v_4/v_2$, we find that $C$ converges to $1$ at very low temperature for $v_4/v_2 > -0.8$, while for $v_4/v_2 \leq -0.8$ it converges towards a smaller value (see Fig.~\ref{fig:crystal} top). This indicates that the lowest energy configurations are no longer the columnar ones. A further indication of the Crystal II phase is given by the average number per site of parallel cubes $\langle N_{\rm cube} \rangle/L^3$ and parallel plaquettes $\langle N_{\rm plaquette}\rangle/L^3$. As expected, we find (see Fig.~\ref{fig:crystal} bottom) that $\langle N_{\rm cube} \rangle/L^3$  decreases from $1/2$ to $0$ and  $\langle N_{\rm plaquette}\rangle /L^3$ from $1$ to $7/12$ quite abruptly as soon as $v_4/v_2 \lesssim -0.8$. Note that the estimate $\left.v_4/v_2\right|_c \simeq -0.8$ that we obtain is quite rough as it is affected by the chosen grid in $v_4/v_2$, the moderate size of the sample ($L=32$) and the finite temperature used in our simulations. Given this, it can be considered as in good agreement with the exact value $-10/12$.

The abrupt behaviour observed in Fig.~\ref{fig:crystal} tends to indicate a first order transition between the Crystal II and columnar phases. This is confirmed by the strong divergence of the specific heat  (Fig.~\ref{fig:result1}), but also by the apparition of a latent heat (see Fig.~\ref{fig:crystal} top). We find that in the full phase diagram, the transition at $T_{c_1}$ is always of first-order nature. When $v_4/v_2 \gtrsim -0.8$, the Crystal II phase and the low-$T$ phase transition disappear. 
  
\subsection{High temperature phase transition $T_{c_2}$} 

The high temperature phase transition corresponds to the simultaneous emergence of dipolar correlations at high temperature and disappearance of the columnar order. 
Before performing a detailed scaling analysis, we already make an important statement : in all simulations for $v_4>0$, we found {\it no evidence} for a first-order phase transition at $T_{c_2}$ between the columnar and the Coulomb phases. This has been checked in all observables at hand (thermodynamics, related to columnar order or the Coulomb phase), including energy histograms at the transition. We will come back to this issue at the end of this section, and in Sec.~\ref{sec:flow}.

\subsubsection{Strongly frustrated regime}

\begin{figure}[h]
 \begin{center}
\includegraphics*[width=7cm]{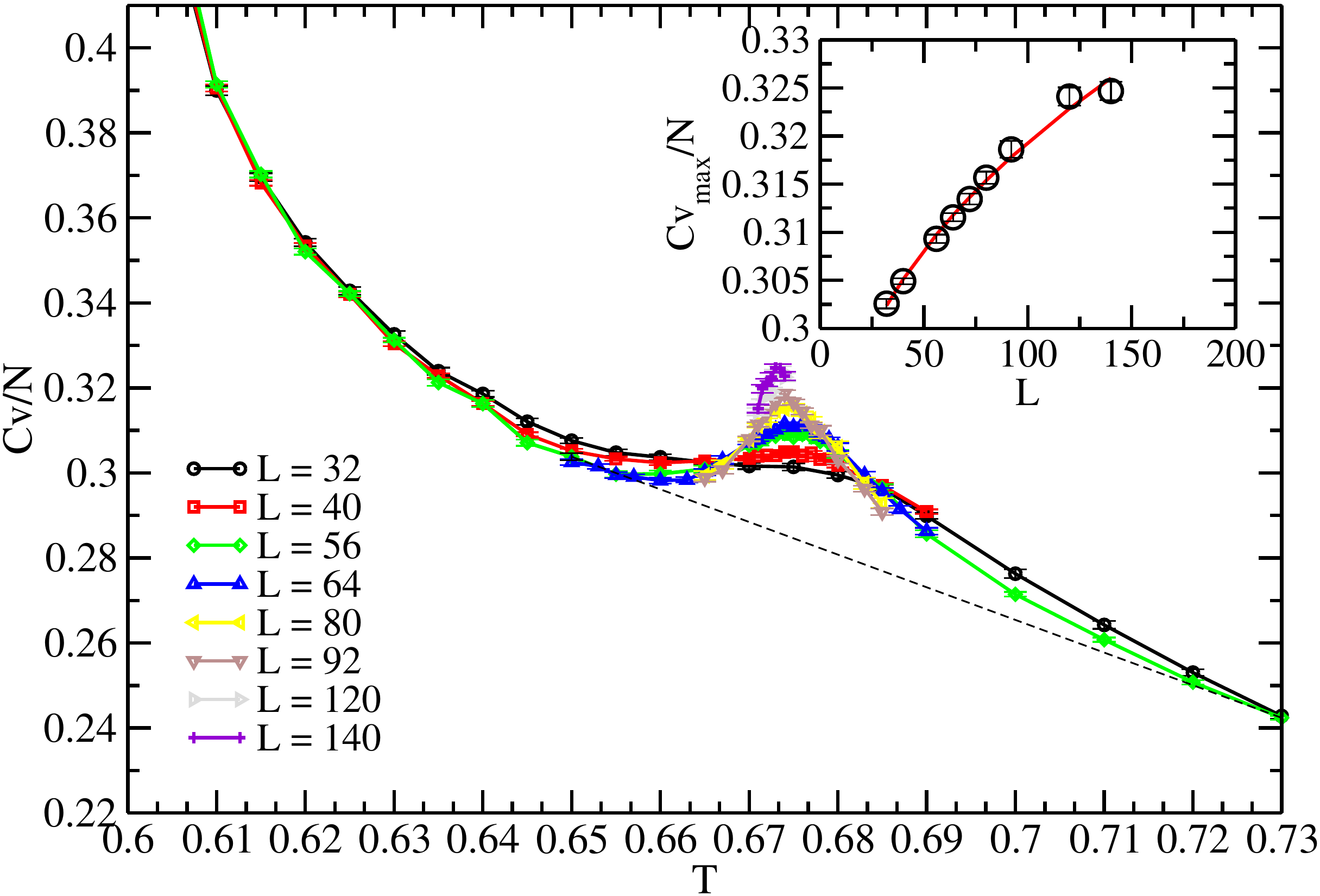}
\includegraphics*[width=7cm]{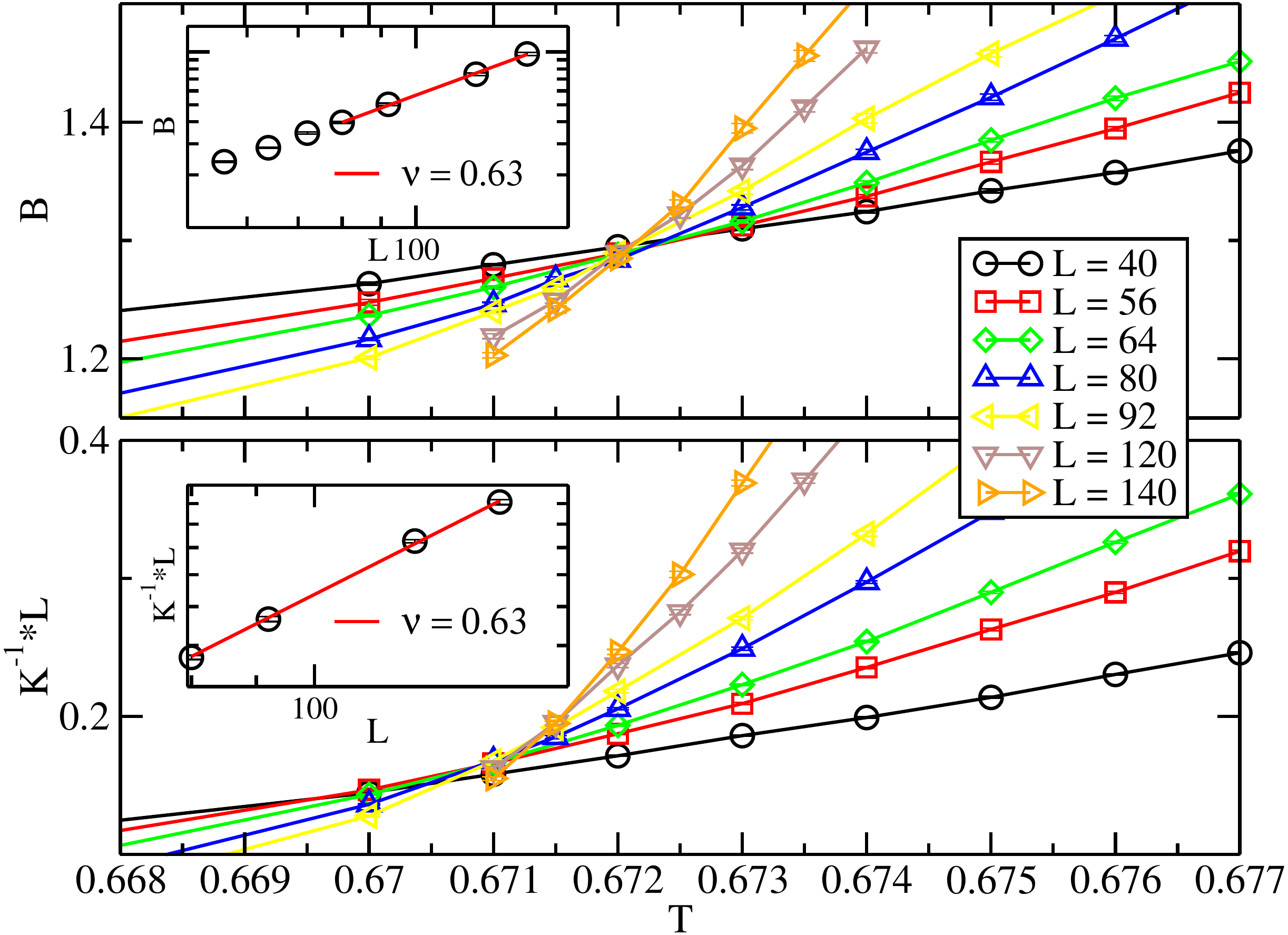}
\includegraphics*[width=7cm]{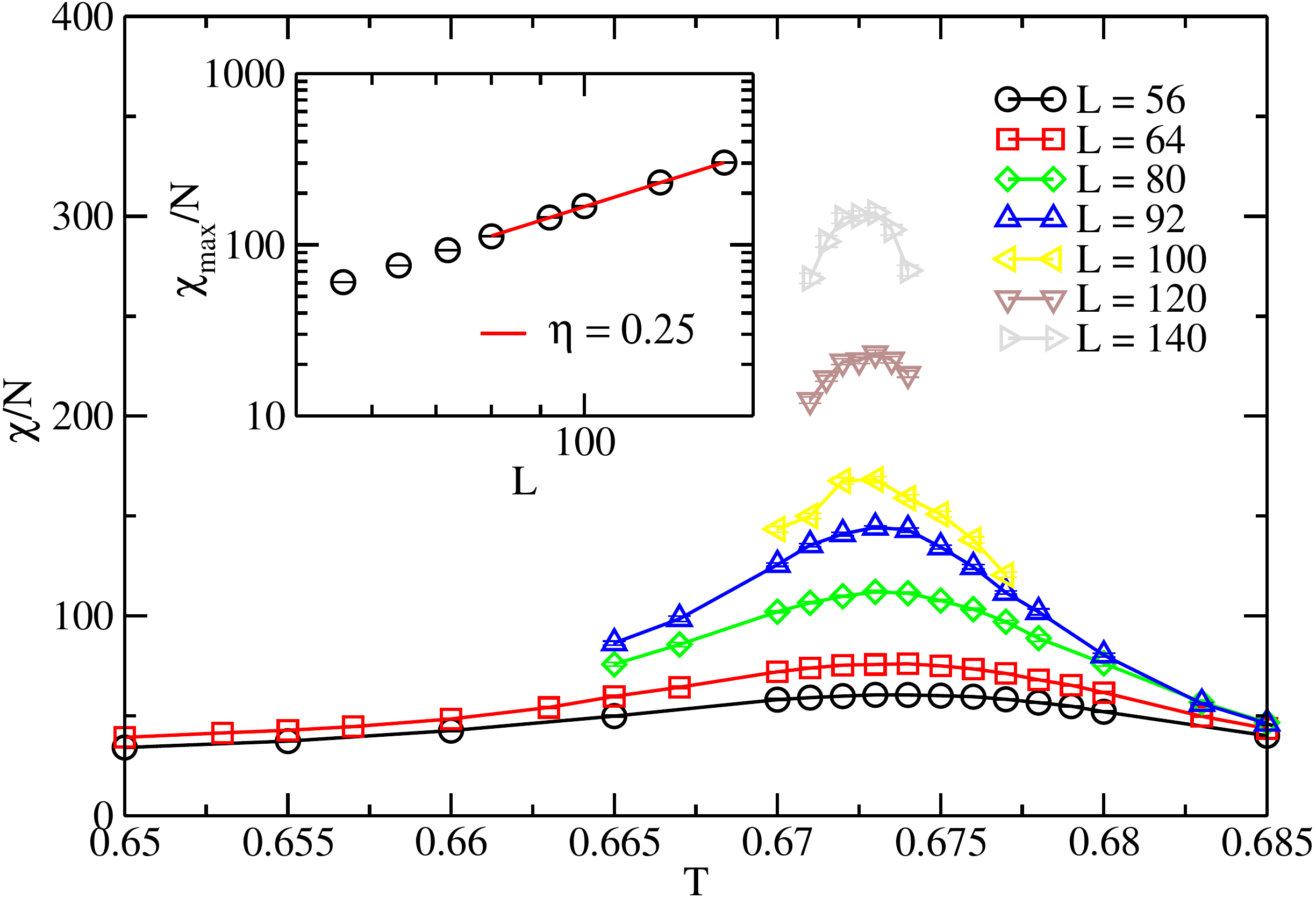}
\caption{ As a function of temperature,  for $v_4/v_2 = -10$ . Top: Specific heat per site $C_v/N$. The dashed line represents a linear interpolation of the regular part of the specific heat. Inset: Maximum of $C_v/N$ as a function of system size.  Middle: Binder cumulant and inverse stiffness  crossings. Insets : Binder cumulant derivative (at $T=0.672$) and derivative $L.dK^{-1}/dT$ (at $T = 0.6715$) as a function of system size (log-log scale). Bottom: Columnar susceptibility $\chi$ per site. Inset: Maximum of $\chi$ as a function of system size (log-log scale). All lines in insets denote power-law fits of critical exponents (see text for details).}
\label{fig:results10}
\end{center}
\end{figure}

In order not to be influenced by any crossover effect, we first concentrate on the transition at $T_{c_2}$ far away from the putative tricritical point at $v_4 = 0$. For $v_4/v_2 = -10$, we performed large scale simulations and applied the FSS analysis to calculate the critical exponents. Our set of data is presented in Fig.~\ref{fig:results10}. Let us first concentrate on the specific heat per site (see Fig.~\ref{fig:results10} top). At the transition, it exhibits a very slow growth with system size, suggestive of a continuous transition. In fact, the divergence is so small that the regular part $C_v^{\rm reg}$ of the specific heat contributes the most even for $L = 140$. The best fit we obtained for system sizes ranging from $L = 32$ to $L=140$ gives an estimate of $\alpha/\nu \sim 0.4$ (Fig.~\ref{fig:results10} top inset). Unfortunately, this estimate varies a lot when considering another subset of system sizes. For instance, discarding the point at $L = 140$ leads to an estimate $\alpha/\nu = 0.6$ and discarding the two points $L=120$ and $L=140$ leads to $\alpha/\nu \sim 0.8$. Thus, while we cannot conclude on the precise value of $\alpha$ at this point, it seems at least to be quite small. The evolution of the Binder cumulant and of the product $K^{-1} . L$ is presented in the middle panel of Fig.~\ref{fig:results10}. For both quantities, we observe a crossing point, in agreement with a second order transition. The two crossing points temperatures are very close: $T_{\rm Binder} = 0.672$ and $T_{\rm Stiffness} = 0.6715$. The thermodynamic measurements of the derivative quantities $dB/dT$ and $L dK^{-1}/dT$, shown in the insets, allows us to have access to two independent estimates of the exponent $\nu$. We find $\nu_{\rm Binder} = \nu_{\rm Stiffness} =  0.63(4)$. These exponents are compatible with the Ising and XY universality classes in $3d$. Moreover, this is consistent with a small value of $\alpha$ assuming hyperscaling $\alpha = 2 - d\nu$. The value of the Binder cumulant and the stiffness at the critical point, $B_c$ and $(K^{-1} \cdot L)_c$, also furnish valuable information, because these quantities are also universal. Checking the crossings of the three largest system sizes of our system, we find $1.28 \leq B_c \leq 1.30$ and $ 0.16 \leq (K^{-1} \cdot L)_c \leq 0.20$. While the value of $B_c$ is consistent with the result obtained for the pure plaquette model $B_c(v_4/v_2 = 0) = 1.27(1)$, the value at criticality for the stiffness is rather smaller $(K^{-1}\cdot L)_c(v_4/v_2 = 0) =0.28(2)$. Finally, we discuss the scaling of the columnar susceptibility (see Fig.~\ref{fig:results10} bottom), obtaining $\eta = 0.25(3)$. The fact that $\eta$ is large and positive for the strongly frustrated system is an unambiguous result of our study, as it is robust for instance on the set of sizes chosen to perform the FSS analysis. Such a strong $\eta$ rules out the possibility of a simple transition such as the Ising or XY universality class. Finally, we note that all the exponents differ considerably from those measured in Ref.~\onlinecite{Alet3d}, implying a different type of transition. This can be already be seen at the qualitative level as  the specific heat does not display any (strong) divergence.

\begin{center}
\begin{table*}
\begin{tabular}{|c|c|c|c|c|c|c|c|}
\hline 
$v_4/v_2$ & $T_c$ & $\alpha/\nu$ & $\nu_{\rm Binder}$ & $\nu_{\rm Stiffness}$ &$\eta$&$B_c$&$(K^{-1}\cdot L)_c $\\
\hline 
-10 & $ 0.672(1) $&$0.4^*$& $0.63(4)$ & $0.63(4)$ & $0.25(3)$ & $1.28 - 1.30$ & $0.16 - 0.20$\\
\hline
-1 & $ 0.953(1) $ &$0.35(10)$& $ 0.60(4) $& $ 0.61(4) $& $0.16(6)$ & $1.27 - 1.29$& $0.18 - 0.22$\\
\hline
-0.2 & $ 1.508(1) $ &$0.80(15)$&$0.50(3)$ &$ 0.58(4)$ & $ -0.02(5)$&$1.26 - 1.28$&$0.23 - 0.27 $\\
\hline
0 & $ 1.675(1) $&$1.11(15) $ & $0.51(3) $& $0.50(4)$ & $-0.02(5) $ & $ 1.26 - 1.28 $ & $0.26-0.29$ \\
\hline
\end{tabular}
\caption{Critical exponents for different frustrating coupling ratios. For the ratio $\alpha/\nu$, the exponents have been measured by a FSS analysis using sizes comprised between $L_{\rm min} = 32$ and $L_{\rm max} = 140$. For the other exponents, we can limit ourselves to larger system sizes between $L_{\rm min} = 80$ and $L_{\rm max} = 140$. In the last row, we remind the results for the pure plaquette model (taken from Ref.~\onlinecite{Alet3d}). The symbol $^*$ denotes one case where the determination of $\alpha/\nu$ is impossible due to the strong contribution of the regular part of the specific heat (see Sec.~\ref{sec:def}).}
\label{tab:tableau2} 
\end{table*}
\end{center}

\subsubsection{Medium and weakly frustrated regime}

\begin{figure}[h]
 \begin{center}
\includegraphics*[width=7cm]{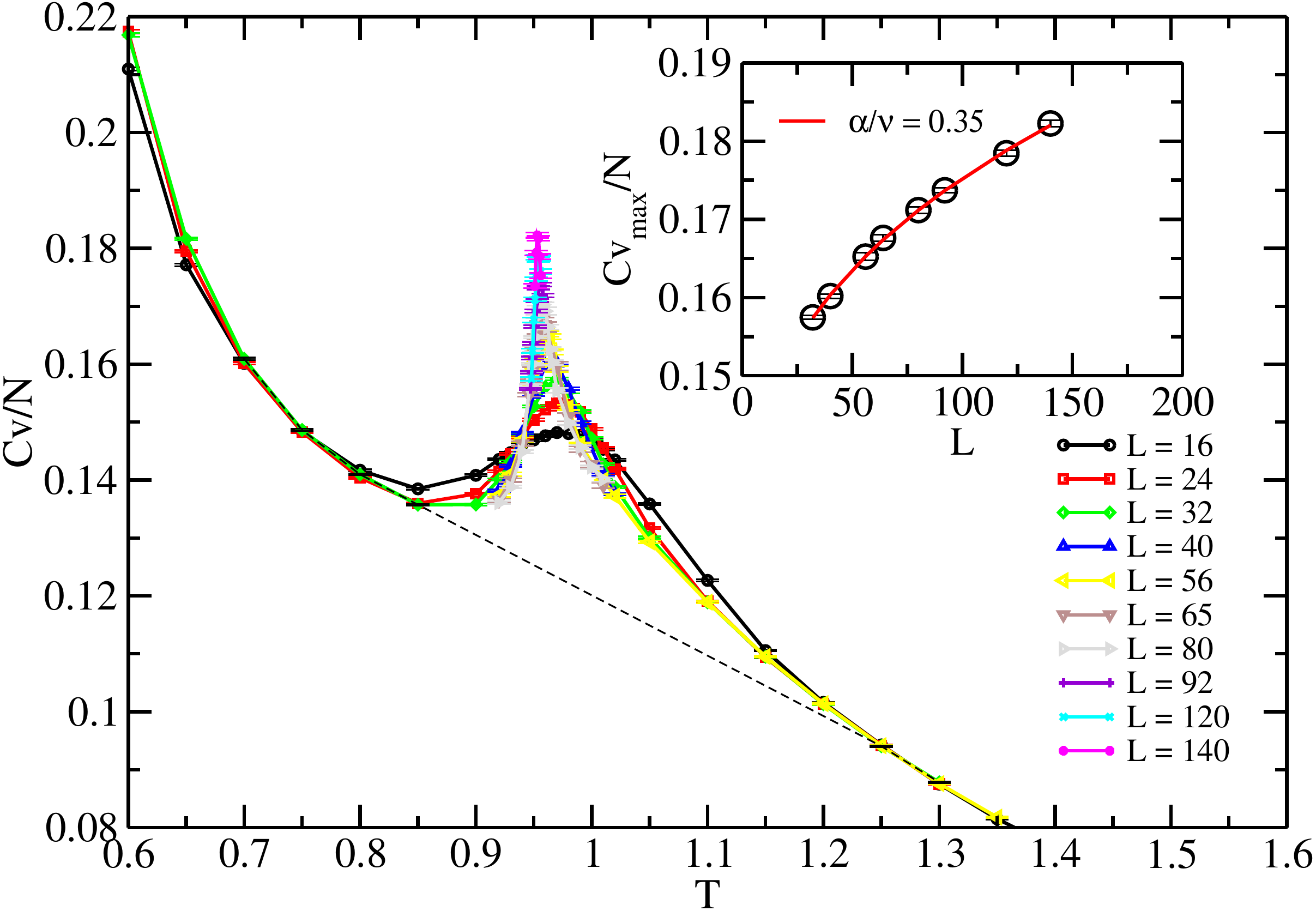}
\caption{ Specific heat per site $C_v/N$ at $v_4/v_2 = -1.0$. The dashed line represents a linear interpolation of the regular part of the specific heat. Inset: Maximum of $C_v/N$ as a function of system size. }
\label{fig:Cv1.0}
\end{center}
\end{figure}

We repeated the same analysis for different values of $v_4/v_2$ on the frustrated side, measuring the exponents $\alpha$, $\nu$ and $\eta$. In particular, we carried out extensive simulations at $v_4/v_2 = -1$ and $v_4/v_2 = -0.2$. Results are summarized in Tab.~\ref{tab:tableau2}. For $v_4/v_2 = -1$, the exponent $\nu$ is compatible with the one obtained for $v_4/v_2 = -10$, and while estimations of $\eta$ are slightly different, the anomalous dimension is clearly non-zero in both cases. The ratio of exponents $\alpha/\nu$ is again the hardest to determine reliably due to the important contribution of the regular part in the specific heat.  For $v_4/v_2 = -1$, $C_v/N$ has however a clear diverging tendency and it is then easier to extract $\alpha/\nu$ (see figure.~\ref{fig:Cv1.0}). We find $\alpha/\nu \sim 0.35$, a value in accordance with the rough estimate at $v4/v2 = -10$ and with the hyper-scaling relation: $\alpha = 2 -d\nu$. 
In any case, the set of exponents we obtain for $v_4/v_2 = -10$ and $v_4/v_2 = -1$ are clearly different from the one obtained for the model with no cubic term. For $v_4/v_2 = -0.2$, the values are on the contrary compatible with the ones obtained from Ref.~\onlinecite{Alet3d}: $\alpha \simeq 0.5$ and $\eta \simeq 0$. Interestingly, we find two non overlapping estimates of $\nu$ from the Binder cumulant and the stiffness derivatives. That may indicate a possible crossover. As a further information, we also give the values of Binder cumulant and stiffness at criticality. 

The results on the frustrated regime bring some interrogations. Our results suggest that there are two different sets of critical exponents (and therefore universality classes) for the Coulomb-columnar phase transition in the extended dimer model: one for the highly frustrated regime with $\alpha \sim 0.2$, $\nu \simeq 0.63$ and $\eta \sim 0.2$ and one close to the point $v_4 = 0$ where the exponents are close to those of the $O(N)$ tricriticality class. Moreover, in the non frustrated regime, we have seen in section \ref{sec:nonfrustrated} that the transition between Coulomb and columnar phases is clearly first order, at least when $v_4/v_2 > 0.4$. A natural interpretation of these data is that there is a tricritical point for a value $v_4^*$ in the vicinity of $v_4 = 0$, separating a continuous transition in the frustrated case to a discontinuous one in the non-frustrated case. The presence of a tricritical point can influence the effective critical exponents measured for values of $v_4/v_2$ in the vicinity of ${v_4^*/v_2}$. Despite the large samples used in our simulations, we are not able to precise the exact location of this tricritical point. In particular, there is no formal reason to believe that the tricritical point is located exactly at $v_4 = 0$. In the next section, we will show that the tricritical point should at least be located on the non frustrated side, that is ${v_4^*/v_2} \geq 0$. 

Another possible explanation that we cannot exclude {\it a priori} is that the transition is always first order. In fact, by looking at the evolution of the different exponents from the non frustrated to the frustrated side, one can perfectly imagine that the correlation length grows continuously but remains \textit{finite} at any $v_4/v_2$. That would mean in particular, that the exponents found in Ref.~\onlinecite{Alet3d} for the pure plaquette model are artifacts caused by a very weakly first order transition. There are several examples of models where reports of unconventional continuous phase transitions have been made and which were finally found to be weakly first order. In order to rule out this scenario, we present in the next section a flowgram analysis combined with a study of histograms at very large system sizes. 

\section{Flowgram analysis}
 \label{sec:flow}

The flowgram method is an advanced version of the FSS analysis which was proposed by Kuklov and coauthors~\cite{Kuklov2}. Consider two points located on the critical line separating the Coulomb and columnar phases in Fig.~\ref{fig:phasediag}. The method relies on the demonstration that the large scale behavior for one given point is identical to that of the second point where the nature of the transition can be easily determined. If this is true, then the two points are in the same critical regime and the nature of the transition remains unchanged all between the two points. On the contrary, a change in the nature of the transition must occur if this is not true. 

The key elements of the method are to {\it (i)} introduce a definition of the critical point for finite-size systems consistent with the thermodynamic limit and which is insensitive to the transition order and {\it (ii)}  compute a quantity $Q$ which is scale invariant at criticality, vanishes in one phase and diverges in the other. To define the operational critical temperature, we assume a finite probability of having a non-zero flux at criticality~\cite{Kuklov1}:
\begin{equation}
\frac{P(\mathbf{\phi} = {\bf 0})}{1- P(\mathbf{\phi} = {\bf 0})} = A,
\end{equation}
where $A$ is some constant which exact value is not relevant for the rest of the method. In practice, we chose A in such a way that it is close to typical values found at the transition for a large system size for $v_4/v_2=-0.2$. We then consider $Q=L.K^{-1}$, as this product is indeed scale invariant for a second order transition, vanishes in the columnar phase (as we expect $K^{-1}$ to vanish exponentally with system size) there and diverges in the Coulomb phase (as $K^{-1}$ is a constant). 

\begin{figure}[h]
 \begin{center}
\includegraphics*[width=7cm]{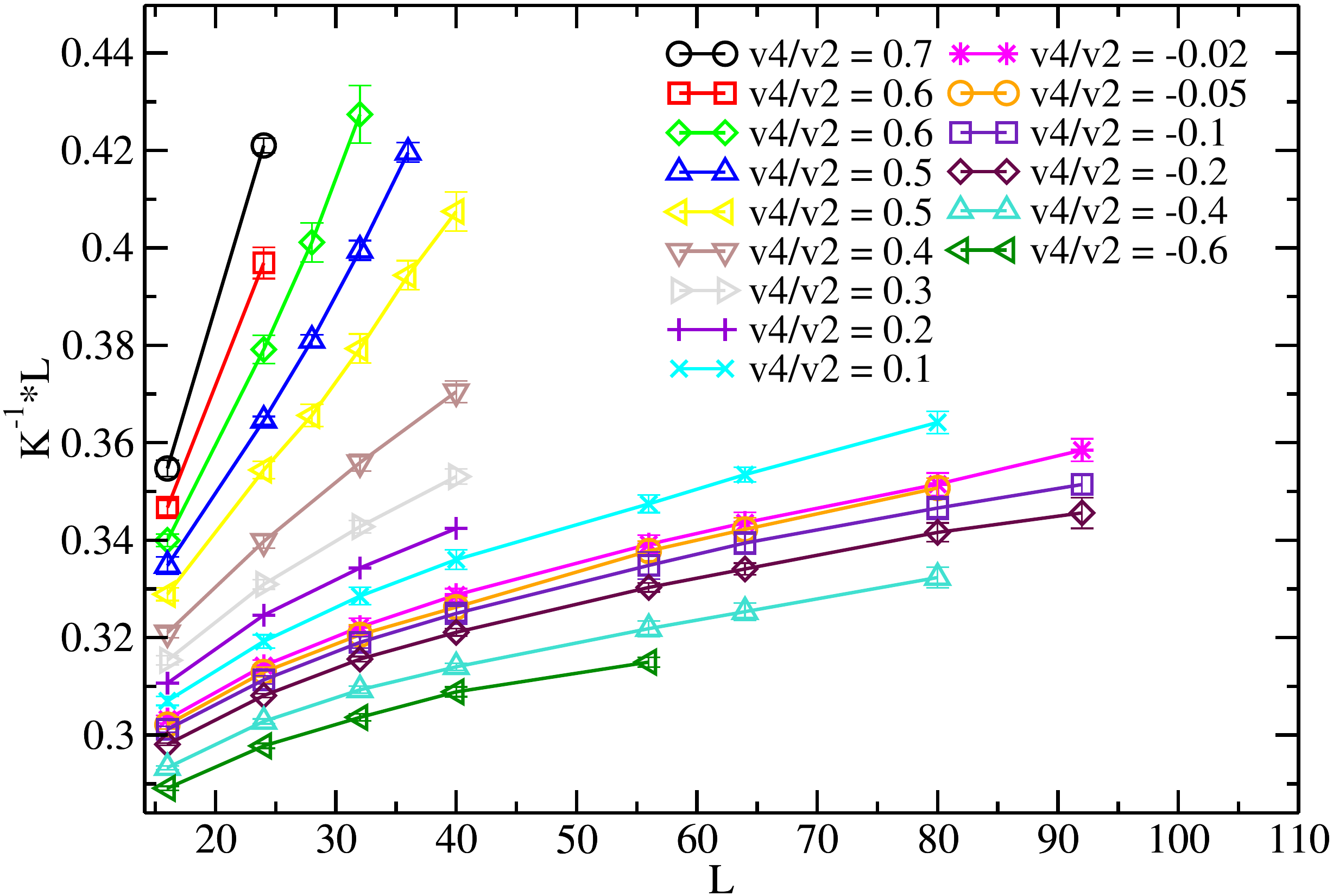}
\includegraphics*[width=7cm]{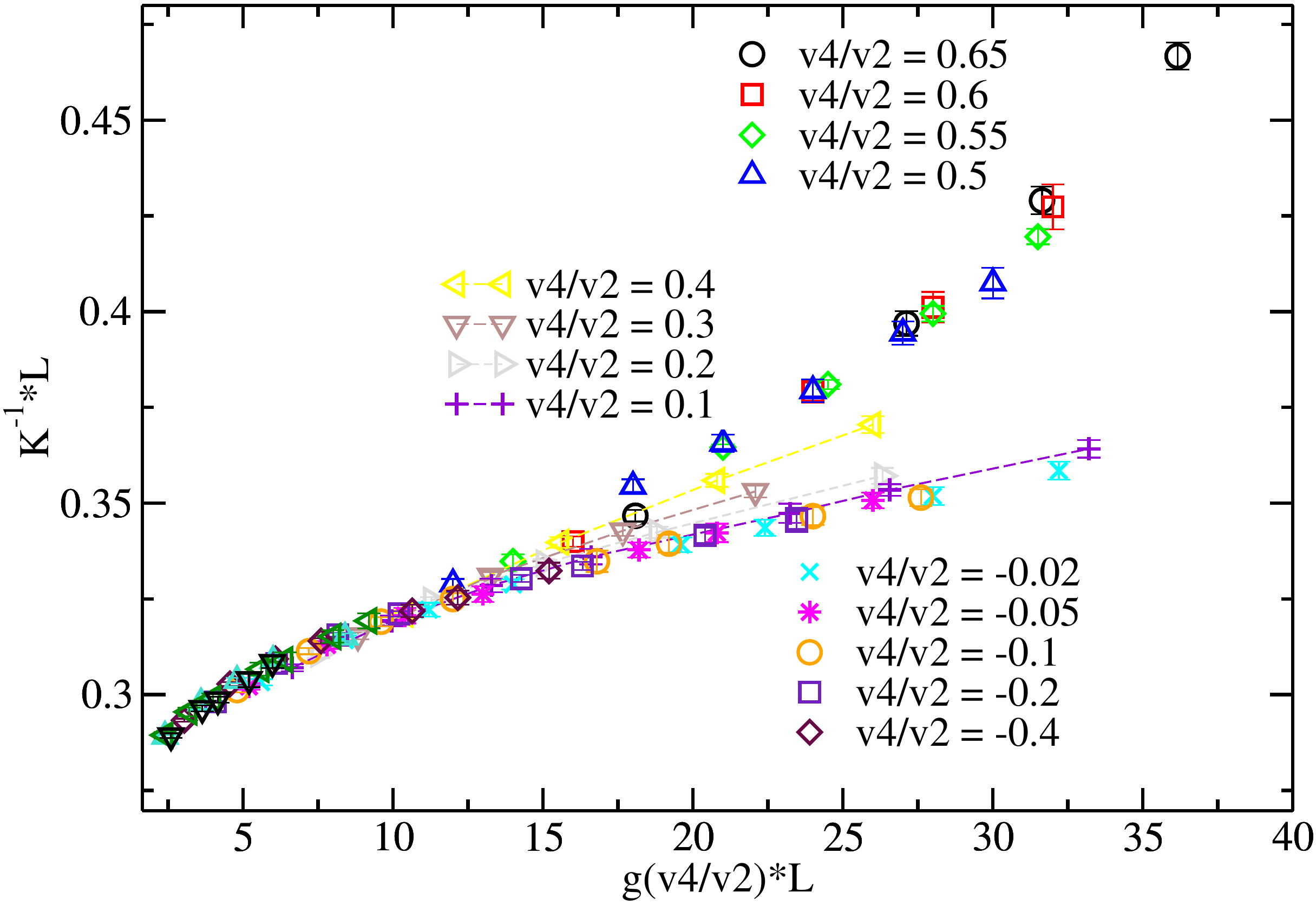}
\caption{Flowgram (top) and performed collapses (bottom) for the extended dimer model. See text for details of the flowgram procedure.}
\label{fig:flowgram}
\end{center}
\end{figure}

The flows $Q_{v_4/v_2}(L)$ for several values of $v_4/v_2$ in the interval $[-0.6,0.7]$ are presented on Fig.~\ref{fig:flowgram} top. The flows can be roughly divided into two groups: for $v_4/v_2 < 0.2$, the flows show a very slow divergence with the system size. For $v_4/v_2 > 0.2$, the flows are strongly diverging. The second group of flows is thus associated with the strongly discontinuous transition. At this stage we cannot draw any conclusion about the first group of curves since it might be that all curves diverge and are actually connected by a scaling transformation. By this, we mean that we need to check if there exists no renormalization fonction $g(v_4/v_2)$ such that plotted as a function of the renormalized length:
\begin{equation}
L_{\rm eff} = g(v_4/v_2) L,
\end{equation}
all the different flows collapse into a single master curve:
\begin{equation}
Q_{v_4/v_2=a}(L_{\rm eff}) = Q_{v_4/v_2=b}(L_{\rm eff}) \; \; \forall \; (a,b).
\end{equation}
 To search for such a transformation, we first try to collapse the flows two by two, starting from the largest positive values of $v_4/v_2$. Fixing $g(v_4/v_2 = 0.7) = 1$, we find the best numerical value for $g(v_4/v_2 = 0.65)$ such that the corresponding flows collapse as a function of $L_{\rm eff}$. We then proceed successively with the next two consecutive values of $v_4/v_2$ and find the best factor for $g(v_4/v_2 = 0.6)$. In this manner, we go through all the parameter space, trying to collapse the curves two by two and finding the corresponding $g(v_4/v_2)$, down to $v_4/v_2 = -0.6$. If such an action is possible, and if the estimated function $g$  varies monotonically as a function of $v_4/v_2$, then all the critical points in the interval $[-0.6,0.7]$  should refer to the same critical behavior. 
 
 We have searched for such a scaling function $g$ and we have arrived to the unambiguous conclusion that it is \textit{not possible} to perform a global collapse of the flows. In particular, the slow divergence of the flow for $v_4/v_2 = -0.02$ is not compatible with the strong growth shown by the flows around $v_4/v_2 = 0.6$. This is because the flows between these two intervals can hardly be collapsed with their neighbors (meaning that for any pair of consecutive flows in the interval $0 < v_4/v_2 < 0.4$, we could not find a rescaling factor such that the two flows are superimposed). On the contrary, we have succeeded in performing two local collapses: one for the region $-0.6 \leq v_4/v_2 \leq 0.02$, and the other for the region $ 0.4 \leq v_4/v_2 \leq 0.7$ (see Fig.~\ref{fig:flowgram} bottom). This is a clear indication of the presence of two different critical behaviors in the phase diagram. Because the collapse for positive $v_4/v_2>0$ is strongly diverging, we naturally associate it with a first order transition. The collapse in the negative range of $v_4/v_2$ describes most probably a second order regime. One can see that this collapse has  actually not converged, as it should for a continuous transition. This is partly due to our original choice of the constant $A$ which made our operational critical temperature slightly above the real $T_c$. A small amount of non-zero stiffness remains present in the system and perturbs the convergence towards the fixed point behavior.
	  
\begin{figure}[h]
 \begin{center}
\includegraphics*[width=8cm]{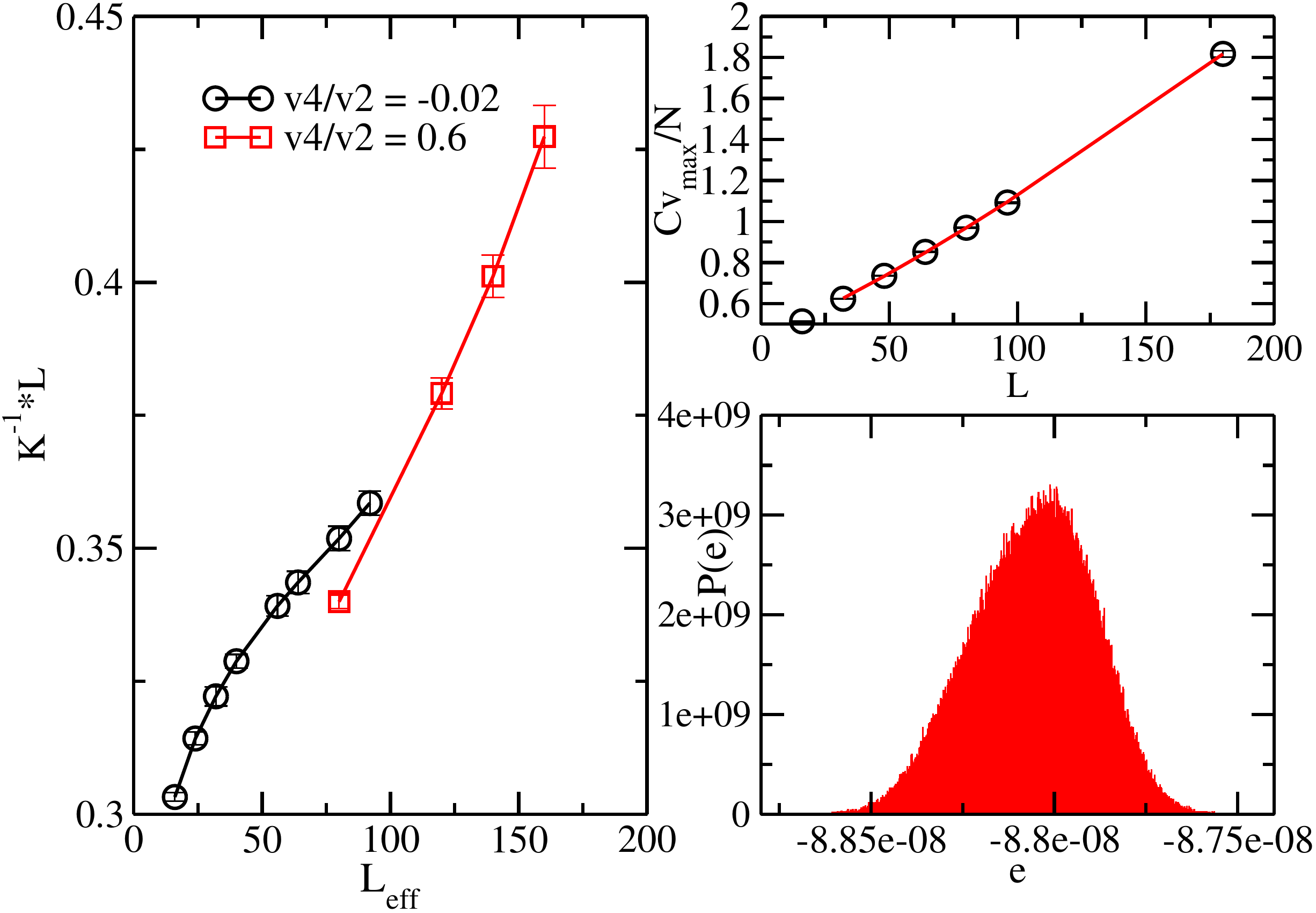}
\caption{Left: Flows at $v_4/v_2 = -0.02$ and $v_4/v_2 = 0.6$ at the rescaled length $L_{\rm eff}$. Top right: Maximum of the specific heat for $v_4 = 0$ as a function of the system size. Data are taken from Ref.~\onlinecite{Alet3d} except for the value at $L = 180$. The solid line is the power-law fit, giving rise to the estimate $\alpha/\nu=1.15(15)$ compatible with the result of Ref.~\onlinecite{Alet3d} {\it without} the knowledge of the L=180 point. Bottom right: Probability of energy per site at $v_4 = 0$ and $L = 180$.}
\label{fig:histo180}
\end{center}
\end{figure}

 To further confirm our conclusion on the flowgram, we have performed the following check. Suppose that the points $v_4/v_2 = -0.02$ and $v_4/v_2 = 0.6$ refer nevertheless to the same critical regime. Then, there should exist a scaling function connecting the flows at $v_4 = -0.02$ and $v_4 = 0.6$. It is in fact possible to connect \textit{very roughly} the two flows by renormalizing the length $L$ for $v_4/v_2 = 0.6$ by a factor $g(0.6) = 8$ while leaving the flow for $v_4/v_2 = -0.02$ unchanged (see Fig.~\ref{fig:histo180} left)).  If this rescaling is really a physical renormalization transformation, then the properties of the system at $v_4/v_2 = -0.02$ and $L = 160$ should correspond to those at $v_4/v_2 = 0.6$ and $L = 32$. For this latest value, we know in particular that the transition is first order since the energy histogram displays a double peak (see Fig.~\ref{fig:histos}). If the critical point $v_4=0$ is indeed in the same regime as the critical point at $v_4/v_2=0.6$, we should expect to see a double-peak also for samples of sizes $L\geq 160$ at $v_4=0$. We have therefore carried a simulation at $v_4 = 0$ with $L = 180$ and have then measured the histogram of energy at the maximum of the specific heat. The histogram displays a unique and well-defined peak (see Fig.~\ref{fig:histo180} bottom right), which confirms that the transition at $v_4=0$ is continuous. Moreover, the value of the specific heat maximum is perfectly compatible with the exponents measured previously at $v_4 = 0$ (see Fig.~\ref{fig:histo180} top right), confirming the validity of the exponent $\alpha / \nu \simeq 1.11(15)$ found in Ref.~\onlinecite{Alet3d}.
 
To conclude on this part, not only can we drive out the possibility that the regimes $v_4/v_2 < 0$ and $v_4/v_2 > 0$ are connected but we can also state that the tricritical point is necessarily located at a value $v_4^*/v_2 \geq 0$. 
   
\section{Discussion}   
 \label{sec:conclusion}
 In this study, we presented an extended version of the classical dimer model with  plaquettes  and cubic interactions. Our simulations indicate that depending on the sign of the interactions, the nature of the critical phase transition between the Coulomb and columnar phases changes. Compared to the pure plaquette model ($v_4/v_2 = 0$), a cubic interaction which reinforces the alinement of dimers ($v_4/v_2 > 0$) leads to a first order transition, identified by a double peak distribution of the energy and a strong divergence in thermodynamic quantities. Whether an infinitesimal positive cubic coupling is enough to alter the nature of the transition is not certain, as we lack any theoretical argument to support this, but simulations tend to indicate that this change should occur very close to $v_4 = 0$. On the other side, when both interactions compete ($v_4/v_2 < 0$), finite size scaling analysis of thermodynamic quantities shows no sign of any discontinuity at the transition. When the cubic interaction is largely dominant ($v_4/v_2 = -10$), the critical exponents deviate significantly from those measured in the pure plaquette model, with $\alpha \sim 0.2$, $\nu \sim 0.6$ and $\eta \sim 0.2$ in the first case and $\alpha \sim 0.5$, $\nu \simeq 0.5$ and $\eta \simeq 0$ in the latter. In the interval $-1 \leq v_4/v_2 \leq 0$, we find exponents in between these two cases, probably due to a cross-over effect. To discharge the possibility of a weak first order transition on the frustrated side, we carried out a flowgram analysis in the vicinity of $v_4 = 0$. The flowgram clearly demonstrates the presence of two groups of flows, one corresponding to positive values of $v_4/v_2$ and the other to negative values. This rules out the possibility of a connection between the two parts of the phase diagram, and thus reveal the presence of a tricritical point~\cite{Kuklov2}. Finally, we also detected the presence of a new crystalline phase at low temperature, deep in the frustrated regime. This phase is characterized by a degeneracy growing with the system size. 
 
In a parallel work, Papanikolaou and Betouras recently studied another extension of the dimer model that slightly differs from ours but which leads to similar conclusions~\cite{Papanikolaou2}. To perturb the pure plaquette model, these authors introduce further neighbor interactions between dimers which preserve the cubic symmetry. They find that when the extra couplings favor the columnar alinement, the transition becomes first order. On the opposite, a frustrating coupling maintains the continuity of the transition. The critical exponents they measure are in agreement with ours in the limit of large frustration. Our study has the advantage of using much larger samples (the maximum system size used in Ref.~\onlinecite{Papanikolaou2} is $L=32$), giving more confidence  on the values of exponents. Furthermore, the use of the flowgram analysis provides an explicit evidence for tricriticality. In another related work, Chen {\it et al.}~\cite{Chen} find that by favoring one particular subset of the 6 columnar orderings at low temperature, the transition between the Coulomb and columnar phases becomes either first order or in the $3d$ $XY$ universality class. 
 
 While not entirely interpretive, it is tempting to compare the exponents on the frustrated side of the dimer model to other models which also display unconventional phase transitions. In Ref.~\onlinecite{Charrier}, a lattice gauge theory representing a coarse-grained version of the dimer model was simulated by Monte Carlo. There, the exponents were found to be very close to the $3d$ $XY$ universality class and thus differ from those presented here. Another source of information is given by the study of the $NCCP^1$ field theory. There exists several indications that the exponent $\eta$ in the $NCCP^1$ theory should be large and positive~\cite{Senthil}. For instance, the transition observed in the ring-exchange models studied in Ref.~\onlinecite{JQ} provide an estimate of $\eta \sim 0.2 - 0.35$ (depending on which correlation function is looked at) and $\nu \sim 0.68(1)$. These values are not far from those we observed in the strong frustration regime. On the other hand, direct simulations of lattice versions of the $NCCP^1$ all exhibit a first order transition~\cite{Kuklov1}. This possibility is ruled out in the dimer model, thanks to our results obtained with the flowgram analysis. \\
 
\section{Acknowledgments}   
We gratefully thank L. Marty for her active participation at an early stage of this work. We also thank S. Trebst for useful discussions. This work was performed using HPC resources from GENCI-CCRT/IDRIS (Grant 2009-100225). We also thank CALMIP for allocation of CPU time. We use the ALPS libraries~\cite{ALPS} for the Monte Carlo simulations.

\end{document}